\documentclass[aps,pra,showpacs,twoside,twocolumn,10pt]{revtex4-1}
\usepackage[colorlinks=true, citecolor=red, urlcolor=blue ]{hyperref}
\usepackage{times,epsfig,amssymb,amsfonts,amsmath,bm,subfigure,mathtools,amsthm,braket,soul,enumitem,color,physics,graphics,graphicx}
\UseRawInputEncoding
\usepackage[normalem]{ulem}
\usepackage{comment}
\usepackage{epstopdf}

\begin{document}

\title{Duality between imperfect resources and measurements for propagating entanglement in networks}
\author{Sudipta Mondal, Pritam Halder, Aditi Sen(De)}
\affiliation{Harish-Chandra Research Institute, A CI of Homi Bhabha National Institute,  Chhatnag Road, Jhunsi, Allahabad - 211019, India}

\begin{abstract}

We propose a measurement-based entanglement propagation strategy for networks in which all nodes except two are initially occupied by a suitably chosen  single-qubit system and the two nodes share a bipartite noisy entangled state. The connections between the sites are established using unsharp two-qubit measurements.  
When only a single node performs measurements, we refer to it as a \textit{unidirectional} protocol while when both parts of the initial entangled states perform measurements, we call it  a \textit{bidirectional}  scheme. When the measurement outcome is post-selected, we demonstrate that in the presence of a local amplitude damping channel acting on a single site, entanglement shareability, as measured by the monogamy score, of the resulting state after measurement can be higher for all values of the strength of the noise than that of the scenario without noise.  We observe that irrespective of the channel, there exists a range of unsharpness parameter where a higher monogamy score may be obtained starting from the initial nonmaximally entangled states than from the initial maximally entangled state. We report that the effect of noise on the average monogamy score entered from the resource state may be reduced faster with the unidirectional protocol than with the bidirectional one. 
\end{abstract}

\maketitle

\section{Introduction}

Over the last century, the remarkable advances in quantum mechanics  have sparked an extraordinary scientific and technological revolution, ushering in a new era of possibilities that transcend the constraints of classical physics. Quantum resources like entanglement \cite{horo_09_ent}, quantum discord \cite{zurek01discord, Bera2018discord}, and coherence \cite{stresltsov_17_coherence}, are essential in the development of quantum appliances. For example, bipartite entangled states shared by  a single sender and  receiver are used to enhance the performance of information processing and communication tasks, e.g., teleportation \cite{Bennett_1993, Bouwmeester_1997, Murao_1999, grudka_2004,Sen(de)_2010}, remote state preparation \cite{Bennett_2005, pati_2000}, dense coding \cite{Bennett_1992, Mattle_1996, Bruss_2004,BRU__2006, de_2011, Horodecki_2012, Shadman_2012,Das_2014}, cryptography \cite{Ekert_1991, Hillery_1999, Shor_2000, Adhikari_2010, Bennett_2014, Sazim_2015, Ray_2016, mudholkar_2023}. In recent years, one of the primary goals has been  to extend these scenarios to encompass arbitrary numbers of parties, resulting in the formation of quantum networks \cite{Kimble2008,wehner_18internet,Pirker2019,Wei2022,azuma2023quantum}, which can either connect several quantum processors or be channels carrying information in quantum communication networks. It has potential applications in the field which  span measurement-based quantum computation \cite{Raussendorf_2001, Hein_2004, Hein_2006, Briegel_2009}, distributed quantum computing \cite{Beals_2013}, quantum secret sharing \cite{Cleve_1999, Karlsson_1999, Jennewein_2000, Gisin_2002, Hillery_2006}, and conference key agreement \cite{Xu_2014} to name a few. The successful implementations of these schemes are possible only when the multiple nodes share a multipartite entangled state which can be created either in a deterministic fashion through gate implementation \cite{Nielsen_2010} or in a probabilistic manner \cite{Walther_2005, Acin_2007, Briegel2009}. Along with theoretical advances, the manifestation of networks is reported in different physical substrates\cite{reiserer22} like photons \cite{pan12},  trapped ions \cite{duan10, Monroe2013}, diamond nitrogen-vacancy centers \cite{Hermans2022}, and quantum dots \cite{Gao2015}. 

To build quantum networks, a few critical points must be addressed in order to achieve scalability of the proposal -- $(1)$ minimum resource required for creating multisite entanglement in networks; $(2)$ optimal operations to be performed; $(3)$ robustness of the protocol against noise or imperfections to be checked. 
The concept of quantum repeaters permits long-distance communication by converting the entire channel into small-distance noisy entangled links that are purified and connected using local operations and classical communication (LOCC) and projective measurements (PM) respectively \cite{briegel_repeater_98,Ladd_2006,Hartmann_2007,Sangouard_2011,Azuma_2015,Pirandola_2016,Azuma_2017,Caleffi_2017,zwerger_18,wehner_18internet,Ratul_2020,Bugalho_2023}. Instead of performing PMs on two nodes, 
 involving more sites in the measurement may result in the creation of multipartite entangled state, which is not plausible in the repeater scenario \cite{Aditi_2005,Cavalcanti_2011,Ratul_2020}. Further,  entanglement transformation via LOCC is also used to connect nodes with entangled states in networks, known as entanglement percolation \cite{Acin_2007,lapeyre09,cuquet09, Broadfoot2009,perseguers10}. In contrast,  unitary circuit-based approaches, such as the generation of cat states by the application of sequential two-qubit controlled-NOT (CNOT) gates on product state of single-qubits \cite{Lee_2005}, fusion techniques \cite{Ozdemir_2011,Zang_2015}, and entanglement circulation protocol with optimized unitary and initial resource corresponding to a particular genuine entanglement content \cite{Pritam_2022} have recently been reported.

Despite the fact that projective measurements can induce genuine multipartite entangled states over long distances and, therefore, networks, the number of initial nodes involved in the protocol always declines, thereby reducing useful links. On the other hand, unsharp measurements \cite{Busch_2013} that can be originated from PMs have the potential to create multipartite entangled network states from an initial state with lesser number of parties, although this avenue still remains almost unexplored. Certain studies have investigated on sequential sharing of quantum correlations, e.g., Bell nonloclaity \cite{Mal_2016}, entanglement \cite{srivastava21}, teleportation \cite{ROY2021127143}, steering \cite{Lv2023}, network nonlocality \cite{Halder_2022,zhang_2023} using unsharp positive operator valued measurements (POVM). Very recently, some of us proposed  creating quantum networks of multipartite entangled states \cite{inflation21} by weak Bell basis measurement \cite{ROY2021127143,srivastava21}.

Proposals for constructing quantum networks with multipartite entangled links often use states and measurements that are not in contact with the environment.
In this work, we provide an architecture for the network which involves imperfections in both resource states to be fed into the network and measurements. In particular, a source transmits a bipartite entangled state into any two nodes of the network via local noisy channels, while all other nodes are occupied by a single qubit state. The links between nodes of an initial state and auxiliary single qubits are created through unsharp measurements. 
 Our goal is to maximize the shareability of entanglement between one of the nodes that shares the initial entangled state and the other nodes in the entangled network.
 The shareability of entanglement is measured with the help of squared negativity  monogamy score \cite{ckw00, ou07, monorev17}, and it is optimized with the parameters of single-qubit system and with unsharpness parameter in the measurement. 
 It is important to note that in networks, noise on resource states can enter only during entanglement sharing, while the number of unsharp measurements might significantly grow when forming links in the network.
We indeed demonstrate that more unsharp measurements can be beneficial in reducing the detrimental impact caused by the noisy channels.
In this general setup, if the unsharp measurement is based on the Bell-basis projective measurement, we determine the recursion relation for obtaining the explicit form of the  resultant state after arbitrary number of rounds, starting from an initial pure state sent through  local noisy channels.

For illustration, we choose three paradigmatic channels, namely amplitude damping channel (ADC), phase damping channel (PDC) and depolarizing channel (DPC) which act on individual sites of the initial pure entangled states. Two distinct scenarios are considered --the behavior of monogamy score is studied when one of the outcomes in the measurement is selected while in the other situation, the averaging is performed over all the measurement outcomes. In the former situation, we demonstrate that  the entanglement shareability can be higher in the presence of noise than in the noiseless case when one of the channel connecting nodes of the network is amplitude damping and that site subsequently performs unsharp measurement, which is above a certain threshold value.
This advantage becomes more pronounced with the increase of number of measurements by the affected nodes. Such counter intuitive  scenario is not present in case of PDC and DPC channels. 
Furthermore, we find that there exist a range of unsharpness parameters, where both post-selected and  the average monogamy score is higher for nonmaximally entangled states than the maximally entangled ones, irrespective of the action of all the local channels. The entanglement shareability is highly disturbed  when one or both the channels are depolarizing. Interestingly, we  observe that more number of unsharp measurements can help to overcome the destructive effects of the noisy channels acted on the resource state.

The paper is structured in the following manner. In Sec. \ref{sec:frame}, we introduce the protocol and its possible varieties which  we refer to as unidirectional and bidirectional protocols, as well as we define the performance indicators of the protocol. We derive the recursion relation for the output state in Sec. \ref{sec:recursion}. By considering a single unsharp measurement and a single auxiliary state, we study the interplay between noise in the initial state and imperfection in the unsharp measurement in Sec. \ref{sec:uni}. The advantage of having ADC and nonmaximally entangled states are discussed in this section also. We perform comparison between the unidirectional and  the bidirectional schemes in Sec. \ref{sec:uni_vs_bi}. In Sec. \ref{sec:conc}, our study concludes with a comprehensive summary of our results.
\section{Framework of networks: Imperfect resource and measurements}
\label{sec:frame}
\begin{figure}
\includegraphics [width=\linewidth]{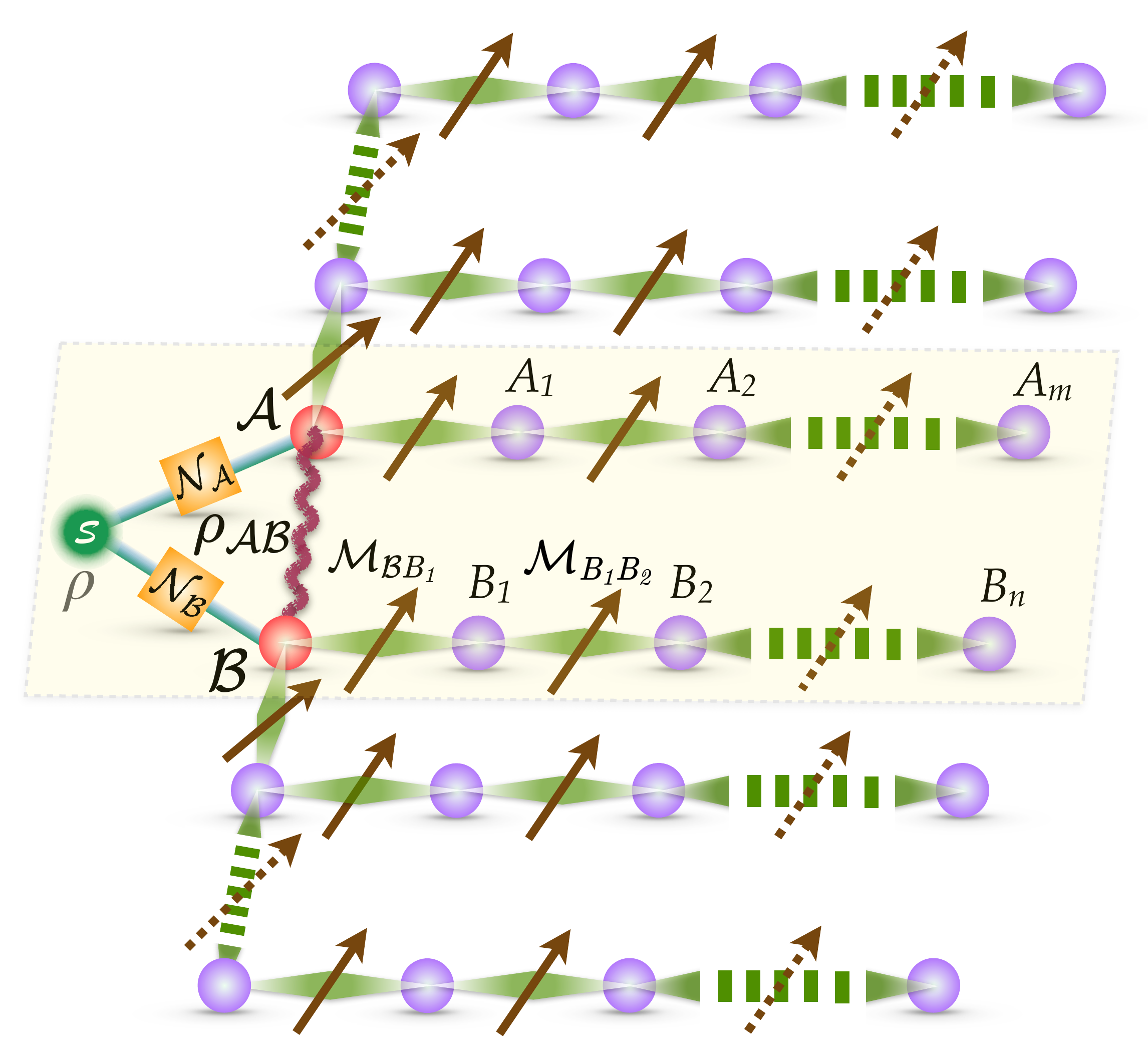}
\caption{Schematic diagram of entanglement propagation protocol for generating  multipartite entangled states. A source, $S$, generates an entangled state, $\rho$, which becomes noisy, $\rho_{\mathcal{AB}}=(\mathcal{N_A}\otimes\mathcal{N_B})\rho$, when shared through local noisy channels, $\mathcal{N_A},\text{ and }\mathcal{N_B}$. 
Other nodes of the networks are occupied by an auxiliary qubits, $\rho_{x_i}$, some are prepared by \(\mathcal{A}\) and \(\mathcal{B}\), denoted by \(A_i\)s and \(B_i\)s respectively. Two-qubit unsharp measurements, $\mathcal{M}$s (green patch) are performed by  nodes \(\mathcal{A}\) and \(\mathcal{B}\) on \(\mathcal{A}A_i\) and \(\mathcal{B}B_i\) respectively, thereby propagating entanglement between nodes. We call a protocol to be an \textit{unidirectional one}, if one of the parties,  say, $\mathcal{B}$ performs unsharp measurement $\mathcal{M}_{\mathcal{B}B_1}$ on the qubits $\mathcal{B}$ and $B_1$ to generate three-qubit entangled state in the first round of the protocol. In a similar way, one can generate $(2+n)$-partite entangled state after the $n$th round of the unidirectional entanglement spreading protocol by conducting measurement on $B_{n-1}$ and $B_n$ auxiliary qubits. On the other hand,  if both $\mathcal{A}$ and $\mathcal{B}$ nodes perform measurements on \(A_1\) and \(B_1\) respectively in the first round and continue similarly in the other rounds, we refer to it as a \textit{bidirectional scheme} (a yellow block). }
\label{fig:schematic}
\end{figure}

Let us propose an entanglement propagation protocol in networks which involves both imperfect measurement and states. The aim is to create multipartite entangled state in which shareability of entanglement among multiple nodes is optimized. The scheme includes two components (see Fig. \ref{fig:schematic}) -- $(1)$ production of an initial resource state and suitable auxiliary single qubits, and $(2)$ the measurement. 

\textbf{$(1)$ Initial resource.} A source, $S$, Producing a bipartite entangled state $\rho$, which is then sent to two nodes of $\mathcal{A}$ and  $\mathcal{B}$ of a network through two quantum channels, $\mathcal{N}_{\mathcal{A}}$ and $\mathcal{N}_{\mathcal{B}}$, which are, in general, noisy. Therefore, an initial resource state, shared by $\mathcal{A}$ and $\mathcal{B}$ can be represented as 
\begin{eqnarray}
    \rho_{\mathcal{A} \mathcal{B}} = (\mathcal{N}_\mathcal{A} \otimes \mathcal{N}_\mathcal{B})\rho.
\end{eqnarray}
Moreover, each local noisy channel affects the state as $\mathcal{N}_{j}(\varrho) = \sum_{i}{K}_i\varrho K_i^{\dagger}$, $j=\mathcal{A}$ and $\mathcal{B}$, where $K_i$ corresponds to the Kraus operators of the given noisy channel with $\sum_{i}K_i^{\dagger}K_i = \mathbb{I}$, $\mathbb{I}$ being the identity operator of the respective Hilbert space. Also, $\mathcal{A}$ and $\mathcal{B}$ possess a collection of auxiliary qubits, $\{\rho_{x_i}=\ket{\chi_i}\bra{\chi_i}\}_{i=1}^k$ which are suitably chosen to spread multipartite entanglement. The sets of auxiliary qubits possessed by $\mathcal{A}$ and $\mathcal{B}$ are denoted respectively by $\mathbf{A}=\{A_i\}_{i=1}^m$, and $\mathbf{B}=\{B_i\}_{i=1}^n$. Here $\ket{\chi_i}=\alpha_i\ket{0}+\beta_i\ket{1}$, where $\alpha_i=\cos{\frac{\theta_i}{2}}, \beta_i=e^{i\phi_i}\sin{\frac{\theta_i}{2}}$ with $\theta_i\in[0,\pi], \phi_i\in[0,2\pi]$ and $k$ denotes the number of auxiliary systems available at each party.

\textbf{$(2)$ Operation for propagation.} To propagate multipartite entanglement in networks, nodes $\mathcal{A}$ and $\mathcal{B}$ have to perform some quantum mechanical operations between the part of an initial shared entangled state and the auxiliary qubit which are capable to produce entanglement as a whole. The joint operation can either be entangling measurement or the two-qubit non-local unitary operator. In the former case, the projective measurement can create entanglement between $\mathcal{A}(\mathcal{B})$ and the auxiliary qubit although it destroys the entanglement between $\mathcal{A}$ and $\mathcal{B}$ and hence such measurement is not suitable to increase the shareability of entanglement in networks. To avoid such situation, two-qubit entangling unsharp measurement (a class of positive operator valued measurement)  characterized by $\mathcal{M}\equiv\{\mathcal{M}^j|j=1,...,4\}$ and $\mathcal{M}^j = M^{j\dagger}M^j$,
satisfying $\sum_{j=1}^4M^{j\dagger}M^{j} = \sum_{j=1}^4 \mathcal{M}^j=\mathbb{I}$, can be applied on $\mathcal{A}(\mathcal{B})$ and $A_{i}(B_{i})$. A prominent example of an entangling, unsharp measurement with a single unsharpness parameter, $0 \leq \lambda \leq 1$, can be written as 
\begin{eqnarray}
    \nonumber \mathcal{M}^j &=& m^j_1(\lambda)\ket{\mathbb{B}^1}\bra{\mathbb{B}^1}+m^j_2(\lambda)\ket{\mathbb{B}^2}\bra{\mathbb{B}^2}\\ &+&m^j_3(\lambda)\ket{\mathbb{B}^3}\bra{\mathbb{B}^3}+m^j_4(\lambda)\ket{\mathbb{B}^4}\bra{\mathbb{B}^4},
    \label{eq:povm_el}
\end{eqnarray}
which is based on the Bell basis, $\big\{\ket{\mathbb{B}^{1(2)}}=\frac{1}{\sqrt{2}}(\ket{00}\pm\ket{11}),\ket{\mathbb{B}^{3(4)}}=\frac{1}{\sqrt{2}}(\ket{10}\pm\ket{01})\big\}$, important for several quantum information processing tasks and $\{m^{j}_k\}_{k=1}^4$ being the probability distribution. Let us consider a situation where information transmission requires spreading of entanglement in the neighbourhood of node $\mathcal{B}$. In that case, to accomplish, say, three-qubit entangled state, $\mathcal{B}$ can prepare an auxiliary system,
$\rho_{B_1}=\ket{\chi_1}\bra{\chi_1}$, and perform unsharp measurement on two qubits, $\mathcal{B}$ and $B_1$, such that output state for the outcome $l_1$ reads as 
\begin{eqnarray}
    \rho^{\{l_1\}}_{\mathcal{A} \mathcal{B} \{B_1\}} = \frac{M_{\mathcal{B}B_1}^{l_1}(\rho_{\mathcal{A} \mathcal{B}}\otimes\rho_{B_1})M_{\mathcal{B}B_1}^{l_1 \dagger}}{\Tr\bigg(M_{\mathcal{B}B_1}^{l_1}(\rho_{\mathcal{A} \mathcal{B}}\otimes\rho_{B_1})M_{\mathcal{B}B_1}^{l_1 \dagger}\bigg)}.
    \label{eq:1st_round_u}
\end{eqnarray}
Therefore, depending on the demand, both $\mathcal{A}$ and $\mathcal{B}$ can perform unsharp measurement by adding auxiliary system in each round of the protocol. 

We refer to the protocol as 
\textit{unidirectional} ($\mathbf{U}$) one if one of the nodes only performs unsharp measurement to spread multipartite entanglement, when  expansion along both the nodes occur, we call it as \textit{bidirectional} ($\mathbf{B}$)  protocol (as shown in Fig. \ref{fig:schematic}). The sharing of multipartite entanglement in this network depends on two crucial factors - one is the (Markovian) noisy quantum channel which is typically responsible for degradation of entanglement in the resource state and the other one depends on the requirement of spreading entanglement around the nodes, $\mathcal{A}$ or $\mathcal{B}$ or both, thereby fixing the nodes on which unsharp measurements have to be performed. Let us explore the different possibilities for multipartite entanglement. Unidirectional protocol with a single noisy channel denoted as $\mathbf{U}(\mathcal{G})_n$ where $\mathcal{G}= \mathcal{A}$ or $\mathcal{B}$ refers to position of the local noise, i.e., one of  the quantum channels $\mathcal{N_A}$ or $\mathcal{N_B}$ during the sharing of the initial entangled state is noisy and $n$ represents the number of auxiliary qubits used or the number of unsharp measurements performed. Let us now illustrate the scenario when node $\mathcal{B}$ is affected by noise. Let us recursively add auxiliary qubits, $B_1$, ...$B_n$ in such a manner that in the $n$th round, adding $\rho_{B_n}$ and executing $\mathcal{M}^n$ on  $B_{n-1}$ and $B_{n}$ with $B_0 = \mathcal{B}$, the output state takes the form as  
\begin{eqnarray}
    \nonumber &&\rho_{\mathcal{A} \mathcal{B}\boldsymbol{B}}^{\boldsymbol{l}} \\ \nonumber &&= \frac{{M}^{l_n}_{B_{n-1} B_n} (\ldots (M^{l_1}_{\mathcal{B} B_1}(\rho_{\mathcal{A} \mathcal{B}}\otimes_{i=1}^n\rho_{B_i})M^{l_1 \dagger}_{\mathcal{B} B_1})\ldots){M}^{l_n \dagger}_{B_{n-1} B_n}}{p^{\boldsymbol{l}}},\\
    \label{eq:nkoutput}
\end{eqnarray}
where $\rho_{\mathcal{A}\mathcal{B}} = (\mathbb{I} \otimes \mathcal{N}_{\mathcal{B}})\rho$,  $\boldsymbol{l}=\{l_1,l_2,\ldots,l_n\}$ denote the outcome of the unsharp measurements performed on $B_iB_j(i\neq j)$ and $p^{\boldsymbol{l}}=\Tr\bigg({M}^{l_n}_{B_{n-1} B_n} (\ldots (M^{l_1}_{\mathcal{B} B_1}(\rho_{\mathcal{A} \mathcal{B}}\otimes_{i=1}^n\rho_{B_i})M^{l_1 \dagger}_{\mathcal{B} B_1})\ldots){M}^{l_n \dagger}_{B_{n-1} B_n})$ is the probability outcome of the string $\boldsymbol{l}$.

Unidirectional scheme with both the channels being noisy \big($\mathbf{U}(\mathcal{A}\mathcal{B})_n$ where $\mathcal{A}\mathcal{B}$ represents the action of both the local channels, $\mathcal{N}_\mathcal{A}$ and $\mathcal{N}_\mathcal{B}$, on $\rho_{\mathcal{A} \mathcal{B}}$\big). In this scenario, $\rho_{\mathcal{A} \mathcal{B}}$ in Eq. (\ref{eq:nkoutput}) can be expressed as $\rho_{\mathcal{A} \mathcal{B}} = (\mathcal{N}_{\mathcal{A}} \otimes \mathcal{N}_{\mathcal{B}})\rho$ and the action of measurement on $\rho_{\mathcal{A} \mathcal{B}}$ can again be represented via Eq. (\ref{eq:nkoutput}).

Bidirectional protocol with a single noisy and two noisy channels, referred to as $\mathbf{B}(\mathcal{G})_n^m$ and $\mathbf{B}(\mathcal{A}\mathcal{B})_n^m$. Here $m$ and $n$ represent the number of auxiliary qubits, possessed by nodes $\mathcal{A}$ and $\mathcal{B}$ respectively. In this situation, the output state after $m$ and $n$ rounds of measurements at sites $\mathcal{A}$ and $\mathcal{B}$ respectively can be  written as
\begin{eqnarray}
   \nonumber &&\rho_{\mathcal{A} \boldsymbol{A} \mathcal{B} \boldsymbol{B}}^{\boldsymbol{k} \boldsymbol{l}} \\ \nonumber &&= \frac{{M}^{k_m}_{A_{m-1} A_m} (\ldots (M^{k_1}_{\mathcal{A} A_1}(\rho_{\mathcal{A} \mathcal{B} \boldsymbol{B}}^{\boldsymbol{l}}\otimes_{j=1}^m\rho_{A_j})M^{k_1 \dagger}_{\mathcal{A} A_1})\ldots){M}^{k_n \dagger}_{A_{m-1} A_m}}{p^{\boldsymbol{k}\boldsymbol{l}}},\\
    \label{eq:klmnoutput}
\end{eqnarray}
where $p^{\boldsymbol{kl}}$ denotes the probability of $\boldsymbol{k} = \{k_1,...k_m\}$ and $\boldsymbol{l}$ strings of outcomes at nodes $\mathcal{A}$ and $\mathcal{B}$ respectively.

\textbf{Assessing the performance of entanglement propagation protocol.} The successful execution of the protocol implies the maximum propagation of entanglement among nodes in the network. In attempt to determine the entanglement content of the resulting state, we choose a  multipartite entanglement measure, $\mathcal{Q}$ which is optimized over the system parameters of the auxiliary states, $\{\theta_i,\phi_i\}_{i=1}^{m(n)}$ and unsharp measurement parameter, $\lambda$ for a fixed initial resource state, i.e.,
\begin{eqnarray}
    \mathcal{Q}^{\boldsymbol{kl}}_{\max}=\max_{\lambda,\theta_i,\phi_i}\mathcal{Q}(\rho^{\boldsymbol{kl}}_{\mathcal{A} \boldsymbol{A} \mathcal{B} \boldsymbol{B}}).
\end{eqnarray}
Note that the above quantity depends on the measurement outcome, i.e., for a fixed outcome strings of one measurement, $\boldsymbol{k}$ and $\boldsymbol{l}$, we set the parameters in auxiliary systems and unsharp measurements in such a way that entanglement in the output state after $(m+n)$ rounds is maximized. To eliminate this outcome dependence, we also perform optimization of the function given by,
\begin{eqnarray}
\mathcal{Q}_{\max}^{\text{avg}}=\max_{\lambda,\theta_i,\phi_i}\sum_{\boldsymbol{kl}}p_{\boldsymbol{kl}}\mathcal{Q}(\rho^{\boldsymbol{kl}}_{\mathcal{A} \boldsymbol{A} \mathcal{B} \boldsymbol{B}})
\label{eq:avg_gme},
\end{eqnarray}
where averaging is performed over all the measurement outcomes and maximization is carried out for the set $\{\theta_i,\phi_i, \lambda\}$.
Both $\mathcal{Q}^{\boldsymbol{kl}}_{\max}$ and $\mathcal{Q}_{\max}^{\text{avg}}$ are functions of initial entanglement in the resource state and noise parameter, p.

\textit{Monogamy Score.} Since we will be dealing with noisy multisite entangled state, computable entanglement  measures are limited in the literature \cite{ggm10,ggm14,tamoghna16mixed_ggm,Buchholz2016}. However, if we are interested how quantum correlations (QC) can be distributed (shared) among many nodes, a natural choice is to adopt the concept of monogamy of entanglement \cite{ckw00,osborne06, Zhang2023, monorev17}. It states that in a multipartite state, if two parties are maximally quantum correlated, other parties cannot share any quantum correlation with these two. Beyond this extreme scenario, it provides an upper bound on the sum of entanglements that a party can share with other sites. Mathematically, for a given multiparty QC measure $\mathcal{Q}$, and a n-party state $\rho_{1 2 \ldots k \ldots n}$ with $k$ being the nodal observer, the monogamy score for $\mathcal{Q}$ reads as \cite{monorev17}
\begin{eqnarray}
    \delta^{\mathcal{Q}}(\rho_{1 \ldots n})= \mathcal{Q}^2(\rho_{k:\text{rest}})-\sum_{i=1,i\neq k}^n \mathcal{Q}^2(\rho_{ki}).
    \label{eq:monogamy}
\end{eqnarray}
where $\rho_{k:\text{rest}}$ and $\rho_{ki}$ represent the multiparty state in the bipartition, $k:\text{rest}$ with rest being  all the parties except party $k$ and the bipartite reduced density matrix between the party $k$ and $i$ respectively. In our work, we choose $\mathcal{Q}$ to be the negativity \cite{vidal02neg,plenio05logneg} which is defined as the sum of the absolute of negative eigenvalues of the partially transposed state \cite{peres96ppt,Horodecki1996}.


\textbf{Interplay between noise content of initial state and unsharpness in measurement.} We also investigate the behavior of entanglement after optimizing only over parameters in auxiliary qubits so that the trade-off relation between unsharpness in measurements and noise in resource state can be examined. As will be argued, in one hand, noise in states is accountable for decreasing entanglement content in networks, and on the other hand, noise in measurement is responsible to create entangled links in networks. We will be concentrating on contrasting nature of these two types of  imperfections associated in this scheme. In the following, we mention four different scenarios of the same.

Having either of $\mathcal{N}_\mathcal{A}$ or $\mathcal{N}_\mathcal{B}$ as a noisy channel, which can be represented by a parameter, say $p$, the state has linear dependence on $p$, and hence the entanglement in networks is affected by $p$, i.e. parameters in $\mathcal{Q}[\mathcal{O}(p)]$ while the  unsharp measurement after the $n$ rounds of the protocol enters the quantity as  $\lambda^n$. Although p and $\lambda^{n}$ enters nonlinearly in the expression of the entanglement for the output state, there is a possibility that positive effects of unsharp measurement dominates over the noisy channels effecting the initial state, thereby leading to a favorable scenario. It can be more pronounced in the case of $\mathbf{B}(\mathcal{G})_n^m$ since in this case, $\mathcal{Q}[\mathcal{O}(p,\lambda_1^m\lambda_2^n)]$ with $\lambda_1$ and $\lambda_2$ being the unsharp parameters of the arm $\mathcal{A}$ and $\mathcal{B}$ respectively as shown in Fig. \ref{fig:schematic}.
In the case where both the channels  $\mathcal{N}_\mathcal{A}$ and $\mathcal{N}_\mathcal{B}$ are noisy,  we can write the influence on  $\mathbf{U}{(\mathcal{A}\mathcal{B})}_n$ is $\mathcal{Q}[\mathcal{O}(p^2,\lambda^n)]$ while it is  $\mathcal{Q}[\mathcal{O}(p^2,\lambda_1^m\lambda_2^n)]$ for $\mathbf{B}(\mathcal{A}\mathcal{B})^m_n$. For small number of nodes present in the network, noise in the resource state can suppress the entanglement propagation although one can expect with the increase of number of nodes, such negative impact can be prevailed in this scheme. As we will illustrate that this is indeed the case although  several counter-intuitive results emerges for networks involving small number of nodes.

\textit{Prototypical noise models.}  In this work we consider three paradigmatic noisy maps, namely, ADC, PDC, DPC whose Kraus operators with noise strength $p$ are given, respectively, by
\begin{eqnarray}
\nonumber &&K^{ADC}_0=
\begin{pmatrix}
1&0\\0&\sqrt{1-p}
\end{pmatrix}, \hspace{0.2cm}
K^{ADC}_1=
\begin{pmatrix}
0&\sqrt{p}\\0&0

\end{pmatrix},\\ \nonumber
&&K_0^{PDC}=\sqrt{1-\frac{p}{2}}\mathbb{I}, \hspace{0.2cm}K_1^{PDC}=\sqrt{\frac{p}{2}}\sigma_z, \\ 
&&K_0^{DPC}=\sqrt{1-\frac{3p}{4}}\mathbb{I},\hspace{0.2cm} K_i^{DPC}=\sqrt{\frac{p}{4}}\sigma_i \forall i\in 1,2,3,
 \label{eq:kraus_adc}
\end{eqnarray}
where $\sigma_1=\sigma_x, \sigma_2=\sigma_y, \sigma_3=\sigma_z$ and $0\leq p\leq1$.

\section{Recursion relation of noisy entanglement shareability (propagation)}
\label{sec:recursion}
Before investigating the trends of multipartite entanglement, it is intriguing to find the exact expression of the resulting state obtained after arbitrary number of measurements, while performed initial state shared via local noisy channels and arbitrary auxiliary qubits. One possible idea is to determine the state after a single measurement which can then be recursively used to obtain the output state at the end of the arbitrary rounds of measurement.

We will now provide a prescription to acquire the recursion relation for the state in this noisy entanglement propagation. To illustrate it, we concentrate on the unidirectional scheme, $\mathbf{U}(\mathcal{B})_n$, and fix the unsharp measurement to be $M^j=\lambda\ket{\mathbb{B}^j}\bra{\mathbb{B}^j}+(1-\lambda)\frac{\mathbb{I}}{4}$, $\{j=1,2,3,4\}$, the initial input state as 
\begin{eqnarray}
    \rho_{\mathcal{A}\mathcal{B}}=(\mathcal{I}\otimes\mathcal{N}_{\mathcal{B}})\ket{\psi}\bra{\psi},
    \label{eq:initial_res}
\end{eqnarray}
with 
\begin{eqnarray}
    \ket{\psi}=\cos{z}\ket{00}+\sin{z}\ket{11}, \hspace{0.2cm} 0\leq z\leq \frac{\pi}{4}.
    \label{eq:istate}
\end{eqnarray}
Here, $\mathcal{N_B}$ being the phase-damping channel, although the approach can similarly be applied to other channels. After the first round of the protocol, the tripartite output state becomes
\begin{eqnarray}
\rho_{\mathcal{A} \mathcal{B}\{B_1\}}^{\{3\}} &=& {\Lambda_1}\rho^{0}_{1} \otimes \ket{\mathbb{B}^1}\bra{\mathbb{B}^1} +{\Lambda_1}\rho^0_2\otimes \ket{\mathbb{B}^1}\bra{\mathbb{B}^2}\nonumber\\
    &+&{\Lambda_2}\rho^0_3\otimes \ket{\mathbb{B}^1}\bra{\mathbb{B}^3}
    +{\Lambda_1}\rho^0_4\otimes \ket{\mathbb{B}^1}\bra{\mathbb{B}^4}\nonumber\\
    &+&{\Lambda_1}\rho^0_5\otimes \ket{\mathbb{B}^2}\bra{\mathbb{B}^1}
    +{\Lambda_1}\rho^0_6\otimes \ket{\mathbb{B}^2}\bra{\mathbb{B}^2}\nonumber\\
    &+&{\Lambda_2}\rho^0_7\otimes \ket{\mathbb{B}^2}\bra{\mathbb{B}^3}
    +{\Lambda_1}\rho^0_8\otimes \ket{\mathbb{B}^2}\bra{\mathbb{B}^4}\nonumber\\
    &+&{\Lambda_2}\rho^0_9\otimes \ket{\mathbb{B}^3}\bra{\mathbb{B}^1}
    +{\Lambda_2}\rho^0_{10}\otimes \ket{\mathbb{B}^3}\bra{\mathbb{B}^2}\nonumber\\
    &+&{\Lambda_3}\rho^0_{11}\otimes \ket{\mathbb{B}^3}\bra{\mathbb{B}^3}
    +{\Lambda_2}\rho^0_{12}\otimes \ket{\mathbb{B}^3}\bra{\mathbb{B}^4}\nonumber\\
    &+&{\Lambda_1}\rho^0_{13}\otimes \ket{\mathbb{B}^4}\bra{\mathbb{B}^1}
    +{\Lambda_1}\rho^0_{14}\otimes \ket{\mathbb{B}^4}\bra{\mathbb{B}^2}\nonumber\\
    &+&{\Lambda_2}\rho^0_{15}\otimes \ket{\mathbb{B}^4}\bra{\mathbb{B}^3}
    +{\Lambda_1}\rho^0_{16}\otimes \ket{\mathbb{B}^4}\bra{\mathbb{B}^4},\nonumber\\
    \label{eq:1st_outcome_state}
\end{eqnarray}
when $M^3$ is the outcome of the measurement. Here, the coefficients, $\Lambda_i$, are functions of $\lambda$, given by
\begin{eqnarray}
    \nonumber \Lambda_1 =\frac{(1-\lambda)}{4},\Lambda_2=\sqrt{\frac{(1+3\lambda)}{4}\frac{(1-\lambda)}{4}},\Lambda_3 = \frac{(1+3\lambda)}{4}.\\
\end{eqnarray}
Again, for the other outcomes, the coefficients in Eq. (\ref{eq:1st_outcome_state}) only change. $\{\rho_i^0\equiv\rho_i^0(z,p,\alpha, \beta)\}_{i=1}^{16}$ are the single-site density matrices expressed as 
\begin{eqnarray}
 \rho^{0}_{1} &=& \varrho_{1}(z,p,\alpha,\beta)^{++++},\rho^{0}_{2} = \varrho_{1}(z,p,\alpha,\beta)^{++--},\nonumber \\ \rho^{0}_{3} &=& \varrho_{2}(z,p,\alpha,\beta)^{++++}, \rho^{0}_{4} = \varrho_{2}(z,p,\alpha,\beta)^{++--}, \nonumber\\ \rho^{0}_{5} &=& \varrho_{1}(z,p,\alpha,\beta)^{+-+-},\rho^{0}_{6} = \varrho_{1}(z,p,\alpha,\beta)^{+--+}, \nonumber\\  \rho^{0}_{7} &=& \varrho_{2}(z,p,\alpha,\beta)^{+-+-}, \rho^{0}_{8} = \varrho_{2}(z,p,\alpha,\beta)^{+--+}, \nonumber\\ \rho^{0}_{9} &=& \varrho_{3}(z,p,\alpha,\beta)^{++++},\rho^{0}_{10} = \varrho_{3}(z,p,\alpha,\beta)^{++--}, \nonumber\\ \rho^{0}_{11} &=& \varrho_{4}(z,p,\alpha,\beta)^{++++}, \rho^{0}_{12} = \varrho_{4}(z,p,\alpha,\beta)^{++--}, \nonumber\\ \rho^{0}_{13} &=& \varrho_{3}(z,p,\alpha,\beta)^{+-+-}, \rho^{0}_{14} = \varrho_{3}(z,p,\alpha,\beta)^{+--+}, \nonumber\\ \nonumber \rho^{0}_{15} &=& \varrho_{4}(z,p,\alpha,\beta)^{+-+-}, \rho^{0}_{16} = \varrho_{4}(z,p,\alpha,\beta)^{+--+},\\
\end{eqnarray}
where the superscripts are used to write them in compact forms. Specifically, they can be represented as 
\begin{eqnarray}
     \nonumber \varrho_{1}(z,p,\alpha,\beta)^{+\pm\pm\pm}&=&\frac{|\alpha|^2}{2} \cos^2{z}\ket{0}\bra{0}\pm\frac{|\beta|^2}{2} \sin^2{z} \ket{1}\bra{1}\nonumber\\ &\pm&(1-\frac{p}{2}) \frac{\beta^*\alpha}{2} \sin{z} \cos{z} \ket{0}\bra{1}\nonumber\\ \nonumber & \pm &(1-\frac{p}{2}) \frac{\beta\alpha^*} {2} \sin{z} \cos{z} \ket{1}\bra{0},\\ \nonumber
     \varrho_{2}(z,p,\alpha,\beta)^{+\pm\pm\pm}&=&\frac{\beta^*\alpha}{2} \cos^2{z}\ket{0}\bra{0} \pm\frac{\beta\alpha^*} {2} \sin^2{z} \ket{1}\bra{1}\nonumber\\ &\pm&(1-\frac{p}{2}) \frac{|\alpha|^2}{2} \sin{z} \cos{z} \ket{0}\bra{1}\nonumber\\\nonumber&\pm& (1-\frac{p}{2}) \frac{|\beta|^2}{2} \sin{z} \cos{z} \ket{1}\bra{0},\\ \nonumber
     \varrho_{3}(z,p,\alpha,\beta)^{+\pm\pm\pm}&=&\frac{\alpha^*\beta}{2} \cos^2{z}\ket{0}\bra{0}\pm\frac{\alpha\beta^*} {2} \sin^2{z} \ket{1}\bra{1}\nonumber\\& \pm &(1-\frac{p}{2}) \frac{|\beta|^2}{2} \sin{z} \cos{z} \ket{0}\bra{1}\nonumber\\\nonumber &\pm & (1-\frac{p}{2}) \frac{|\alpha|^2}{2} \sin{z} \cos{z} \ket{1}\bra{0},
\end{eqnarray} 
and
\begin{eqnarray}
      \nonumber \varrho_{4}(z,p,\alpha,\beta)^{+\pm\pm\pm}&=&\frac{|\beta|^2}{2} \cos^2{z}\ket{0}\bra{0}\pm\frac{|\alpha|^2}{2} \sin^2{z} \ket{1}\bra{1}\nonumber\\ &\pm&(1-\frac{p}{2}) \frac{\alpha^*\beta}{2} \sin{z} \cos{z} \ket{0}\bra{1}\nonumber\\ &\pm &(1-\frac{p}{2}) \frac{\alpha\beta^*} {2} \sin{z} \cos{z} \ket{1}\bra{0}.
\end{eqnarray}
Similarly, we can find out the output state in the first round of the protocol for an arbitrary outcome of the measurement. After performing $n$ measurements with the starting node being $\mathcal{B}$, when $l_n$ outcome is obtained, the resulting state reads as 
\begin{eqnarray}
   \rho_{\mathcal{A} \mathcal{B}\boldsymbol{B}}^{\boldsymbol{l}} = \sum_{i=1}^4\sum_{r=(4i-3)}^{4i}c_{ir}^{l_n}\rho_r^n(z,p,\alpha,\beta)\otimes\ket{\mathbb{B}^i}\bra{\mathbb{B}^r},
   \label{eq:recursion}
\end{eqnarray}
with coefficients $c_{ir}^{l_n}=\sqrt{m^{l_n}_i m^{l_n}_r}$ where $m^{l_n}_j$s are coefficients of unsharp measurement elements in Eq. (\ref{eq:povm_el}). Moreover,
\begin{eqnarray}
    \nonumber \rho_1^n(z,p,\alpha,\beta) &=& \sum_{i=1}^4\sum_{r=(4i-3)}^{4i}c_{ir}^{l_{n-1}}\rho_r^{n-1}(z,p,\alpha,\beta)\\ \nonumber&&\otimes \rho_r^0(\pi/4,0,\alpha,\beta),\\ \nonumber 
    \rho_{i_1}^n(z,p,\alpha,\beta)&=& (\sum_{l=1,5,9,13}P_{l,l+1}-\sum_{l=3,7,11,15}P_{l,l+1}){\rho^n_{i_1-1}}\\ \nonumber &&\forall i_1=2x:1\leq x\leq8,
\end{eqnarray}
\begin{eqnarray}
    \nonumber \rho_{i_2}^n(z,p,\alpha,\beta)&=&\sum_{l=1}^2(P_{l,l+2}+P_{l+4,l+6}+P_{l+8,l+10}\\ \nonumber&+&P_{l+12,l+14}){\rho^n_{{i_2}-2}}\forall i_2\in 3,7,11,15,\\ \nonumber
    \rho_{i_3}^n(z,p,\alpha,\beta)&=&\sum_{l=1}^4(P_{l,l+4}-P_{l+8,l+12})\rho^n_{i_3-4}\forall i_3\in 5,13.\\ 
    \text{and }\rho_9^n &=& \sum_{l=1}^8 P_{l,l+8}\rho_1^n.
\end{eqnarray}
Here, $P_{r,s}$ denotes the swap operator which has the effect on $\rho_r^n$ as shown below:
\begin{eqnarray}
    \nonumber P_{r,s}&&(c_{ir}^{l_{n-1}}\rho_r^{n-1}(z,p)\otimes \rho_r^0(\pi/4,0,\alpha,\beta)\\ \nonumber&+&c_{is}^{l_{n-1}}\rho_s^{n-1}(z,p)\otimes \rho_s^0(\pi/4,0,\alpha,\beta))\\ \nonumber &\rightarrow&(c_{ir}^{l_{n-1}}\rho_r^{n-1}(z,p)\otimes \rho_s^0(\pi/4,0,\alpha,\beta)\\ \nonumber&+&c_{is}^{l_{n-1}}\rho_s^{n-1}(z,p)\otimes \rho_r^0(\pi/4,0,\alpha,\beta)).
\end{eqnarray}

Similar prescription can be used to write down the recursion relation for any noise model on the resource state and also for the bidirectional case which involve modifications of $\Lambda$ and the coefficients, $\rho_j^0$s. In fact, the form of the recursion relation in Eq. (\ref{eq:recursion}) remains unchanged except the expression of $\rho_j^0$s which depend on the form of the initial resource state and the noise parameter of the channel.

\section{Unidirectional entanglement propagation: Noise in Source vs. unsharpness in Measurement}
\label{sec:uni}
Let us first set the stage to address the question of duality that we expect from the noisy initial state and weak measurements. We will first deal with the simplest scenario involving three parties, i.e., one of the parties, $\mathcal{A}$ or $\mathcal{B}$ performs an unsharp measurement on the part of the initial state and auxiliary qubit. In this set up, there are two noisy scenarios --
$(i)$ either  $\mathcal{N}_{\mathcal{A}}$ or  $\mathcal{N}_{\mathcal{B}}$ is noisy) and $(ii)$ $\mathcal{N}_{\mathcal{A}}$ and  $\mathcal{N}_{\mathcal{B}}$ both are noisy. In both the  situations, monogamy score is computed with $\mathcal{B}$ being taken as the nodal observer, since unsharp measurement is carried out on parties $\mathcal{B}$ and  ${B}_1$ which are responsible to spread entanglement in networks. 

Typically, noisy channels suppress the entanglement content of the state. However, we will demonstrate that unsharp measurement induces the capability of sharing entanglement in networks even in presence of noise. As mentioned before, we will compute both post-selected as well as averaged squared negativity monogamy score, denoted by $\delta_{\mathcal{B}}^{\{k\}}$ and  $\delta^{\text{avg}}_{\mathcal{B}}$. Since we take squared negativity as an entanglement measure during computation of monogamy score, we omit $N^2$ from the notation of monogamy score in Eq. (\ref{eq:monogamy}) and subscript denotes the nodal site.


Let us first concentrate on the post-selection, i.e., we study the features of the output state when the specific outcome is occurred. In particular, we are interested to analyze the sharing capability of the entanglement content for the resulting state, which we obtain by maximizing the monogamy score, $\delta_{\mathcal{B}}^{\{3\}}$ over the auxiliary system (parameterized by $\theta_1$, $\phi_1$ ) and by varying unsharp parameter $\lambda$.

Initially, we deal with the local amplitude damping channel, $\mathcal{N}_\mathcal{B}$, which acts on the $\mathcal{B}$ node of the initial pure state, $\ket{\psi}$. The unsharp measurement is performed on $\mathcal{B}$ and ${B}_1$ which can result to an output state, revealing interesting features in shareability. Let us highlight them. 

\begin{figure}
\includegraphics [width=\linewidth]{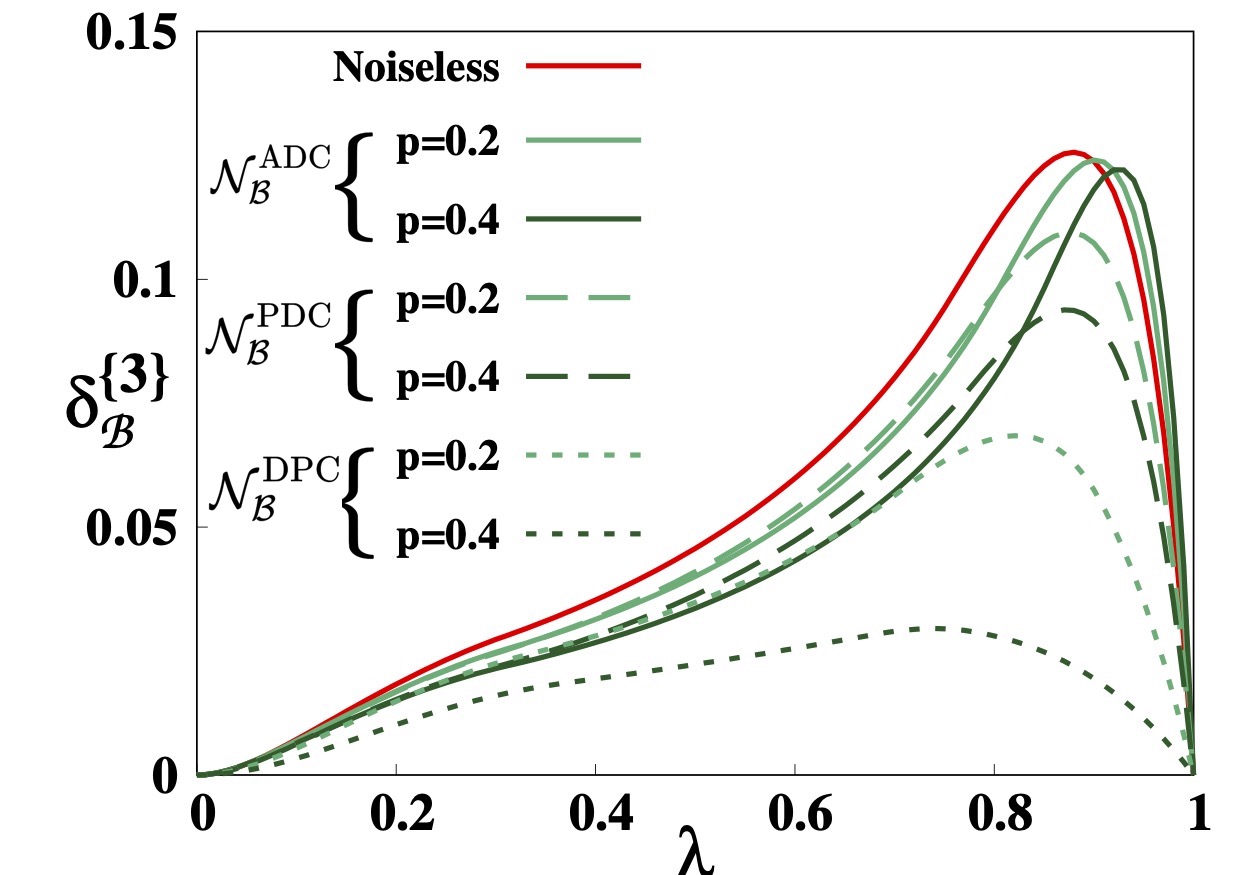}
\caption{Monogamy score, $\delta_{\mathcal{B}}^{\{3\}}$ (ordinate) of the tripartite entangled state generated from  the post-selection of the measurement outcome, \(M^3\) in the unidirectional case, $\mathbf{U}(\mathcal{B})_1$  against the unsharp parameter, $\lambda$ (abscissa) in Eq. (\ref{eq:povm_el}). Local noise in the form of ADC (solid green), PDC (dashed green), DPC (dotted green) with noise strength $p=0.2 \text{ (light-green)}$, $0.4\text{ (dark-green)}$  act on $\mathcal{B}$ of the initial resource state, \(|\psi\rangle = \cos z |00 \rangle + \sin z |11\rangle\),  having \(z = \pi/8\). Red solid line corresponds to the noiseless scenario. In presence of local ADC,  higher monogamy score  is obtained than that in the noiseless case  for some higher values of \(\lambda\). Such observation remains true even when the unidirectional protocol is carried out involving the  higher number of parties. 
Both the axes are dimensionless.}
\label{fig:u1_case}
\end{figure}

$1.$  First of all, $\delta_{\mathcal{B}}^{\{k\}}(\rho_{\mathcal{A}\mathcal{B}\{\text{B}_1\}},p,\lambda)$ (with the auxiliary string, \(\boldsymbol{B}\) containing only a single site $B_1$) increases with the increase of $\lambda$ till $\lambda_{\text{max}}$ at which it reaches maximum for a fixed noise strength $p$ and an initial state parameter $z$ as shown in Fig. \ref{fig:u1_case}.
When $\lambda > \lambda_{\text{max}}$, it starts decreasing before vanishing at $\lambda =1$. 

$2.$ To highlight its counter intuitive behavior, we define a quantity called critical unsharpness present in measurement, denoted as $\lambda_{c}$ where 
\begin{eqnarray}
    \nonumber \delta_{\mathcal{B}}^{\{k\}}(\rho_{\mathcal{A}\mathcal{B}\{\text{B}_1\}},p=0)\bigg|_{\lambda = \lambda_c}=\delta_{\mathcal{B}}^{\{k\}}(\rho_{\mathcal{A}\mathcal{B}\{\text{B}_1\}},p\neq 0)\bigg|_{\lambda = \lambda_c},\\
    \label{eq:lambda_c}
\end{eqnarray}
i.e., after post-selecting the measurement result, $k$, monogamy scores coincide for states affected by noise and without noise. 

Interestingly, in a tripartite scenario, we observe that there exist a region with $\lambda > \lambda_{c}$, in which $\delta_{\mathcal{B}}^{\{3\}}\big(\rho_{\mathcal{A}\mathcal{B}\{\text{B}_1\}},p\neq 0,\lambda\big) >\delta_{\mathcal{B}}^{3}\big(\rho_{\mathcal{A}\mathcal{B}\{\text{B}_1\}},p =0, \lambda \big)$ with $z\leq \frac{\pi}{4}$ in the shared state $\ket \psi$, i.e., noisy states can produce an output state which has potential to share more entanglement in networks than that of the final state which is not affected by noise. It possibly indicates the complementary relation between imperfection in measurements and states which also supports dimensional arguments presented in the previous section.
We call this phenomena as \textit{constructive impact on entanglement sharing capability}. Such positive influence is possible only when the noise acted on the part of the  state and the unsharp measurement involving the party coincide. Specifically, to obtain such counter intuitive result, we require unsharpness parameters, to be very high, so that measurement is close to a projective measurement although local noise in the state can be as high as possible.  

$3.$ When $\lambda > \lambda_{c}$ and in presence of optimal $p$, the optimal auxiliary system turns out to be $\ket 0 \bra 0$ (i.e., $\theta_{1}=0$ and independent of $\phi_{1}$). The expression for monogamy score reduces in this region as 
\begin{widetext}
    \begin{eqnarray}
    \delta_{\mathcal{B}}^{\{3\}}(\rho_{\mathcal{A}\mathcal{B}\{B_1\}},p,\lambda)&=&
\bigg\{-{\bigg(4 {\Lambda_{1} } (p+1)-\frac{1}{2} \sqrt{64 \lambda ^2 (p-1)^2 \sin ^4{z}+4 (4 {\Lambda_{1} })^2 (-(p-1) \cos {2 z}+p+1)^2}+4 {\Lambda_{1} } (1-p) \cos {2 z}\bigg)^2}\nonumber\\&+&4 \bigg(\sqrt{\left(-2 \lambda  p^2+p^2+\lambda ^2 (p (5 p-8)+4)\right) \sin ^4{z}+8 {\Lambda_1} (1-p) (\lambda +4  {\Lambda_2 }+1) \sin ^2{z} \cos ^2{z}}-4 {\Lambda_1} p \sin ^2{z}\bigg)^2 \nonumber\\&-&\bigg(R_{1}-\sqrt{2} \sqrt{\sin ^4{z} R_{2}+(\lambda -1) (p-1) (\lambda +4 {\Lambda_2 }+1) \sin ^2{2 z}}\bigg)^2\bigg\}/R_{3} \nonumber\\
\end{eqnarray}
\end{widetext}
 where 
\begin{eqnarray}
    R_{1}&=&\sin^{2}z (\lambda -4 {\Lambda_2 }-3 \lambda  p+4 {\Lambda_2 } p+p+1),\nonumber\\R_{2}&=&(-\lambda  (\lambda +4 {\Lambda_2 }-2)-4 {\Lambda_2}\nonumber\\&+&p^2 (\lambda  (3 \lambda -3\ 4 {\Lambda_2 }-2)+4 {\Lambda_2 }+1)\nonumber\\&+&4\lambda  (4 {\Lambda_2 }-1) p+1),\nonumber\\\text{and }R_{3}&=&(-4 \lambda  p+4 \lambda  (p-1) \cos 2z +4)^2.
\end{eqnarray}

$4.$ The benefit of unsharp measurement can also be observed, when for any nonvanishing  noise strength, $p_{1} > p_{2}$ in the ADC, we obtain $\delta_{\mathcal{B}}^{\{3\}}\big(\rho_{\mathcal{A}\mathcal{B}\{\text{B}_1\}},p_{1},\lambda\big) >\delta_{\mathcal{B}}^{\{3\}}\big(\rho_{\mathcal{A}\mathcal{B}\{\text{B}_1\}},p_{2}, \lambda \big)$ with $z \leq \frac{\pi}{4}$ and when both $0 \leq p_{1},p_{2} < 1$. Note that such scenario does not occur when $z > \frac{\pi}{4}$, which is a clear indication of the fact that this phenomena does not only depend on the initial entanglement of the resource state but also on the structure of the state. Importantly, the effect of ADC on a qubit has a biasness towards $\ket{0}$ on the Bloch sphere. Therefore, initial states with different structures but same entanglement give rise to different features in case of ADC.

$5.$ \textit{Haar-uniform resource states.}  Let us now choose  Haar-uniformly generated two-qubit pure states \cite{Bengtsson2006} of the form, $\ket{\psi^r} = a_1 \ket{00}  + a_2 \ket{01}+ a_3 \ket{10} + a_4\ket{11}$ with $a_i=a'_i+ia''_i$, as the initial resource states of our protocol. Here, $a'_i,a''_i$ are real numbers, chosen randomly from the Gaussian distribution with vanishing mean and unit standard deviation. We perform numerical simulation over $5\times10^5$ such states, among which $20\%$ states shows that having ADC acting on it in the node $\mathcal{B}$ generates higher monogamy score than that of noiseless scenario.

$\textbf{Remark 1:}$ Similar positive impact of ADC, i.e., constructive impact on entanglement sharing capability persists even when compute monogamy score with respect $\mathcal{A}$, i.e.,  $\delta_{\mathcal{A}}^{\{3\}}$.

$\textbf{Remark 2:}$ When noise acts on the $\mathcal{A}$ part or both the parts of the shared state locally, and the monogamy is considered with respect to $\mathcal{B}$,  no such advantage is found.

Furthermore, the results obtained for ADC do not hold for the phase damping and the DPC. Although, some common features emerge in the monogamy score for all the local channels (see Fig. \ref{fig:u1_case}). Firstly, for a fixed noise strength, $p$, in the channel, there always exists $\lambda_{\text{max}}$ at which $\delta_{\mathcal{B}}^{\{3\}}$ achieves its maximum value. $\lambda_{\text{max}}$ depends both on the channel, $\mathcal{N}$ and the strength $p$, thereby providing different $\lambda_{\text{max}}$ for different $\mathcal{N}$ and $p$ values. However, when the initial shared pure state possess a fixed amount of initial entanglement, $\lambda_{\text{max}}$ lies in the same neighborhood of $\lambda$ which is independent of the channels (comparing maximum in Fig. \ref{fig:u1_case}). Secondly, $\delta_{\mathcal{B}}^{\{3\}}\big(\rho_{\mathcal{A}\mathcal{B}\{\text{B}_1\}},p_{1},\lambda\big) >\delta_{\mathcal{B}}^{\{3\}}\big(\rho_{\mathcal{A}\mathcal{B}\{\text{B}_1\}},p_{2}, \lambda \big)$, when $p_{1} < p_{2}$ for all values of $\lambda$ for all channels except ADC with $\lambda > \lambda_{c}$ (see Fig. \ref{fig:u1_case}).

\begin{figure}
\includegraphics [width=\linewidth]{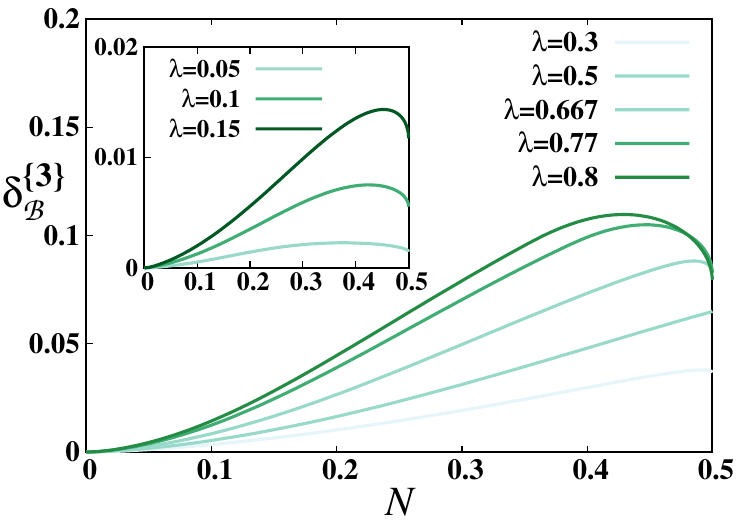}
\caption{Monogamy score (post-selected), $\delta_{\mathcal{B}}^{\{3\}}$  (vertical axis) vs  negativity of the initial pure state, \(|\psi\rangle\), ${N}$ (horizontal axis). Different curves represent different  values of the unsharp parameter, $\lambda$ for a fixed value of  the noise strength in the ADC, $p=0.2$ acted on \(\mathcal{B}\) only. In case of  the unidirectional protocol,  for a fixed value of \(\lambda\),  nonmaximally entangled pure states perform better  than the maximally entangled ones except certain range of $\lambda$ which will be  illustrated in Fig. \ref{fig:neg_lam_state_opt}. The vertical and horizontal axis of the inset are the same as in the main figure. Only the small \(\lambda\) values are considered in the inset. Both the  axis are dimensionless.}
\label{fig:initial_ent_case}
\end{figure}

\textbf{Gain in entanglement propagation by nonmaximally entangled initial state.} During designing of entanglement propagating protocol, it is crucial to determine the  requirement for the initial resource which is also important in establishing its robustness. The protocol presented here has three layers by which entanglement either is inserted in networks or gets affected  --  (i) the source produces a bipartite entangled state to be inserted in networks as  the initial resource, (ii) the initial entanglement gets disturbed after the action of the noisy channel, and (iii) finally, the entanglement is shared among three parties after performing the entangling unsharp measurement. Let us concentrate on the first one, i.e., the dependence of the initial shared entanglement quantified with negativity, $N$, on $\delta_{\mathcal{B}}^{\{3\}}$ of the final state after post-selection. For a fixed amount of noise strength in the channel and unsharpness in measurements, i.e., for fixed $p$ and $\lambda$, we observe that there are two states with $N
(\ket {\psi^{1}_{\mathcal{A}\mathcal{B}}}) < N
(\ket {\psi^{2}_{\mathcal{A}\mathcal{B}}})$  for which $\delta_{\mathcal{B}}^{\{3\}}\big(\rho^{1}_{\mathcal{A}\mathcal{B}\{\text{B}_1\}},p,\lambda\big) >\delta_{\mathcal{B}}^{\{3\}}\big(\rho^{2}_{\mathcal{A}\mathcal{B}\{\text{B}_1\}},p, \lambda \big)$ occurs where $\rho^{i}_{\mathcal{A}\mathcal{B}\{\text{B}_1\}}$ is obtained from the initial pure state, $\ket {\psi^{i}} $ ($i= 1,2$) as depicted in Fig. \ref{fig:initial_ent_case}.
Further, we find that such a benefit from the nonmaximally entangled states can be obtained when the entanglement is high enough to overcome the action of noise on the state with the aid of unsharp measurement. This result again confirms that the effect of noise on the entanglement in the initial state can be nullified  with the assistance of entanglement present in the unsharp measurement. 
\begin{figure}
\includegraphics [width=\linewidth]{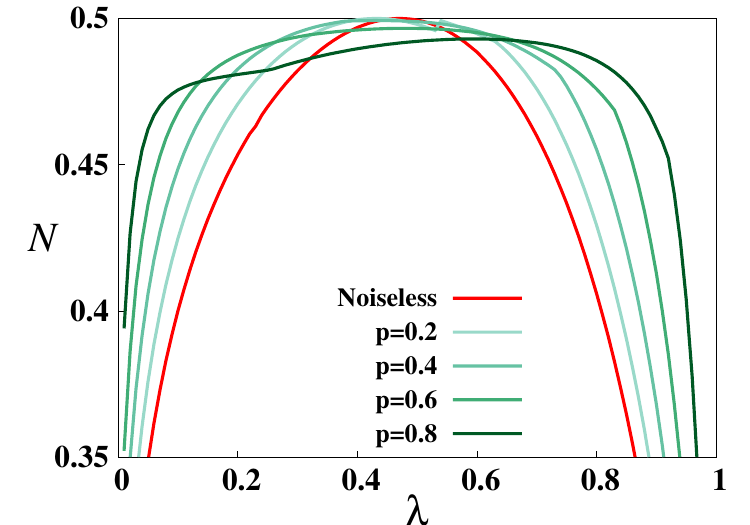}
\caption{Negativity, ${N}$ (ordinate) of optimal initial resource state  against control parameter, $\lambda$ (abscissa), of the unsharp measurement for different fixed values of  local ADC, with $p=0.2,0.4,0.6$ and ,$0.8$ (sequentially light green to dark green) and $p=0$ (noiseless case) $\text{(red)}$. Maximally entangled state is the optimal resource in the range of unsharp parameter, $\lambda=0.44 $ to $ 0.5$ for the noiseless scenario. This window shifts to a lower value of  $\lambda=0.39$ to $0.47$ when $p=0.2$. As the noise strength increases, maximally entangled states  no longer remains optimal. Both the axes are dimensionless.}  
\label{fig:neg_lam_state_opt}
\end{figure}
Fig. \ref{fig:neg_lam_state_opt} shows that the observation is not monotonic with the unsharpness amount, $\lambda$. Specifically, both low and high values of $\lambda$ provide advantage in propagating entanglement for nonmaximally entangled states possessing moderate amount of entanglement which is no more true for moderate values of $\lambda$, i.e., there exists a region of $\lambda$ where the maximally entangled state is the most suitable input to be shared, leading  to the optimal shareability of entanglement.

Let us probe the situation towards illustrating the role of entanglement in the initial resource state. To do this, for a fixed $p$ and $\lambda$, we calculate the optimal initial entangled state which maximizes the monogamy score of the output state $\rho^{\{3\}}_{\mathcal{AB}{\{B_1\}}}$, i.e., we evaluate 
\begin{eqnarray}
    \delta^{\{3\}}(\rho'_{\mathcal{AB}\{B_1\}},p,\lambda)=\max_{\theta_1,\phi_1,\mathcal{E}(\ket{\psi})}\delta^{\{3\}}(\rho_{\mathcal{AB}\{B_1\}},p,\lambda),
\end{eqnarray}
where along with maximization of auxiliary qubit, we perform optimization over the initial entanglement, $\mathcal{E}(\ket{\psi'})$ which leads to a maximum $\delta^{\{3\}}(\rho'_{\mathcal{AB}\{B_1\}},p,\lambda)$ for a fixed $p$ and $\lambda$ of a given channel. See Fig. \ref{fig:neg_lam_state_opt} for ADC. Firstly, we notice that in the noiseless case, for the shared maximum entanglement content of $\ket{\psi}$, there exists a range of $\lambda$ which leads to a maximum generation of entanglement in the resulting state. In other words, for each values of $\lambda$, there exists a fixed initial entangled pure state which gives maximum $\delta^{\{3\}}(\rho_{\mathcal{AB}\{B_1\}},p=0,\lambda)$. Secondly, both in case of high and low value of $\lambda$, nonmaximally entangled states are superior than the maximally entangled shared state. Interestingly, with the increase of local noise acted on $\mathcal{B}$, the maximum $N$ required in resource state which achieves  $\max_{\theta_1,\phi_1,\mathcal{E}(\ket{\psi})}\delta^{\{3\}}(\rho_{\mathcal{AB}\{B_1\}},p,\lambda)$ decreases although the range of $\lambda$ increases by which maximum monogamy score can be accomplished. The output state for PDC and DPC exhibit similar behavior which further confirm the existence of nonmaximally entangled states as optimized input states.

\textbf{Comparison among channels: ADC vs PDC vs DPC.} We have shown that the constructive effect of unsharpness present in the measurement can overcome the noise affecting the resource state when one of the outcomes are selected. Let us investigate the average response of measurement on noisy states during entanglement propagation, when one of the exemplary noisy channels acts either on $\mathcal{A}$, or $\mathcal{B}$ or $\mathcal{A}$ and $\mathcal{B}$. Moreover, averaging over the outcomes in the measurement makes the presence of noise in the channels more prominent which can also make the comparison between the channels more meaningful.
Let us compute $\delta^{\text{avg}}_{\mathcal{B}}$ after optimizing over the parameters involved in the auxiliary qubits, so that $\delta^{\text{avg}}_{\mathcal{B}}$ is a function of initial resource state, $z$ as well as both noise strengths, $p$ and unsharpness  $\lambda$. It will be interesting to probe one after another.

{\it Imperfect measurements vs noisy states.}
Let us first fix $p$ and vary $\lambda$ with a fixed initial entanglement although the results remain qualitatively same if one varies $z$. The observations can be enumerated as follow. 

$1.$ There is an overall decrease in monogamy score compared to the noiseless situation for all values of the unsharp parameter regardless of any form of the noise present in the channels.

\begin{figure}
\includegraphics [width=\linewidth]{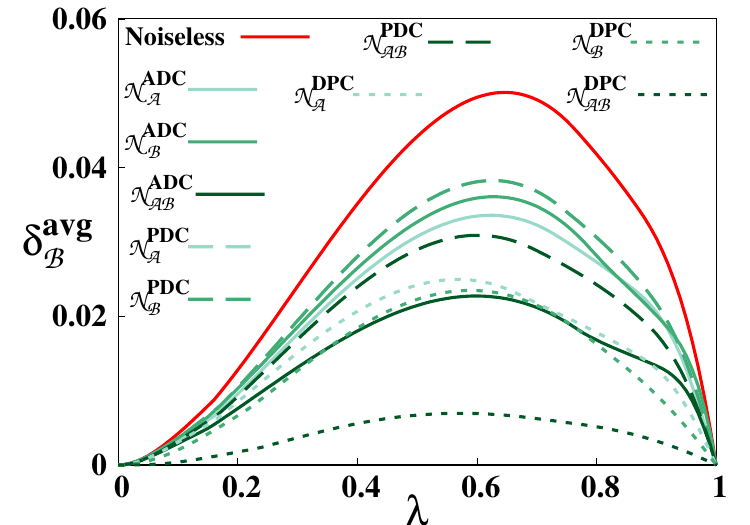}
\caption{Average monogamy score, $\delta_{\mathcal{B}}^{\text{avg}}$ ($y$-axis) of the tripartite entangled state with respect to unsharp parameter, $\lambda$ ($x$-axis) for a fixed value of local noise,  $p=0.4$. Here the averaging is performed over the measurement outcomes of the unsharp measurements, given in Eq. (\ref{eq:povm_el}). Noise acts either on \(\mathcal{A}\) or \(\mathcal{B}\) or on both, denoting them as  $\mathcal{N}_{\mathcal{A}}$, $\mathcal{N}_{\mathcal{B}}$, $\mathcal{N}_{\mathcal{A}\mathcal{B}}$ (sequentially light green to dark green) in the form of ADC (solid), PDC (dashed), DPC (dotted) and $p=0$ represents the noiseless case (red solid line). Here  the initial resource state  $\ket {\psi}$  is with $z=\frac{\pi}{8}$. Both the axes are dimensionless.}
\label{fig:avg_mono_lam}
\end{figure}

$2.$ Clearly, when local channels act on both the parties, $\delta^{\text{avg}}_{\mathcal{B}}$ is lower than that for the case of a single noisy channel. In general, we observe their ordering for all values of $\lambda$ as 
\begin{eqnarray}
    \delta^{\text{avg}}_{\mathcal{B}} \big(\mathcal{N}^{\text{ADC}}_{\mathcal{B}} (\rho_{\mathcal{A}\mathcal{B}\{{B}_1\}})\big) &>& \delta^{\text{avg}}_{\mathcal{B}} \big(\mathcal{N}^{\text{ADC}}_{\mathcal{A}} (\rho_{\mathcal{A}\mathcal{B}\{{B}_1\}})\big)\nonumber\\ &>& \delta^{\text{avg}}_{\mathcal{B}} \big(\mathcal{N}^{\text{ADC}}_{\mathcal{A}\mathcal{B}} (\rho_{\mathcal{A}\mathcal{B}\{{B}_1\}})\big),\nonumber
\end{eqnarray}
\begin{eqnarray}
    \delta^{\text{avg}}_{\mathcal{B}} \big(\mathcal{N}^{\text{DPC}}_{\mathcal{A}} (\rho_{\mathcal{A}\mathcal{B}\{{B}_1\}})\big) &>& \delta^{\text{avg}}_{\mathcal{B}} \big(\mathcal{N}^{\text{DPC}}_{\mathcal{B}} (\rho_{\mathcal{A}\mathcal{B}\{{B}_1\}})\big)\nonumber\\ &>& \delta^{\text{avg}}_{\mathcal{B}} \big(\mathcal{N}^{\text{DPC}}_{\mathcal{A}\mathcal{B}} (\rho_{\mathcal{A}\mathcal{B}\{{B}_1\}})\big),\nonumber
\end{eqnarray}
\begin{eqnarray}
    \nonumber\text{and }\delta^{\text{avg}}_{\mathcal{B}} \big(\mathcal{N}^{\text{PDC}}_{\mathcal{A}} (\rho_{\mathcal{A}\mathcal{B}\{{B}_1\}})\big) &=& \delta^{\text{avg}}_{\mathcal{B}} \big(\mathcal{N}^{\text{PDC}}_{\mathcal{A}} (\rho_{\mathcal{A}\mathcal{B}\{{B}_1\}})\big)\nonumber\\ \nonumber &>& \delta^{\text{avg}}_{\mathcal{B}} \big(\mathcal{N}^{\text{PDC}}_{\mathcal{A}\mathcal{B}} (\rho_{\mathcal{A}\mathcal{B}\{{B}_1\}})\big),\\
\end{eqnarray}where we assume that both the local channels in $\mathcal{N}_{\mathcal{A}\mathcal{B}}=\mathcal{N_A}\otimes\mathcal{N_B}$ 
 have the same strengths in noise, $p$, as can be matched  from Fig. \ref{fig:avg_mono_lam}.
 Notice that, in case of PDC acted on any one of the qubit of $\ket{\psi}$, the noisy input state takes the form as 
 \begin{eqnarray}
    \rho_{\mathcal{A}\mathcal{B}} &=& \cos^{2}{z} \ket{00} \bra{00} + \big(1-\frac{p}{2}\big) \sin{z} \cos{z} \ket{11}\bra{00} \nonumber\\ \nonumber&&+ \big(1-\frac{p}{2}\big) \sin{z} \cos{z} \ket{00} \bra{11} + \sin^2{z} \ket{11}\bra{11},
\end{eqnarray}
which causes the output monogamy score to have the same value when noisy channel acts one of the sites of \(|\psi\rangle\). Our analysis suggests that for different values of unsharp parameter and noise content, the difference in the monogamy scores between the independent actions of single noisy channels, i.e., either $\mathcal{N}_\mathcal{A}$ or $\mathcal{N}_\mathcal{B}$ can be made arbitrarily small than that of $\mathcal{N}_{\mathcal{A}\mathcal{B}}$. Therefore, in case of unidirectional three-party output, we can possibly write
$\delta^{\text{avg}}_{\mathcal{B}} \big(\mathcal{N}_{\mathcal{B}} (\rho_{\mathcal{A}\mathcal{B}\{{B}_1\}})\big) \approx \delta^{\text{avg}}_{\mathcal{B}} \big(\mathcal{N}_{\mathcal{A}} (\rho_{\mathcal{A}\mathcal{B}\{{B}_1\}})\big)> \delta^{\text{avg}}_{\mathcal{B}} \big(\mathcal{N}_{\mathcal{A}\mathcal{B}} (\rho_{\mathcal{A}\mathcal{B}\{{B}_1\}})\big)$.

\begin{figure}
\includegraphics [width=\linewidth]{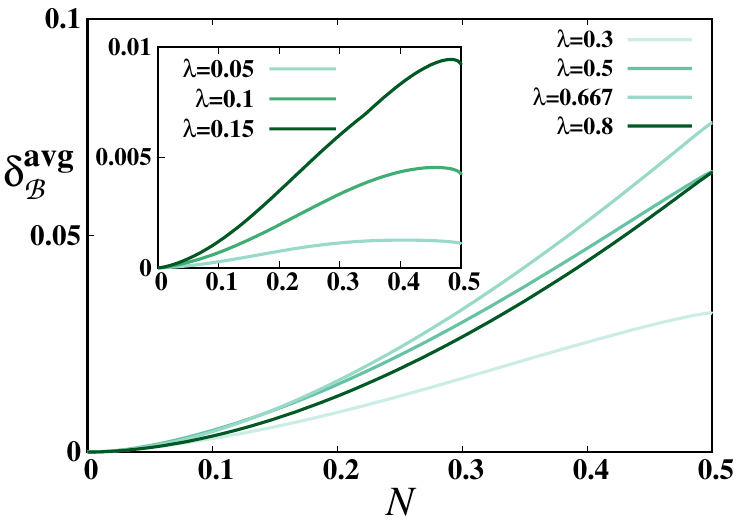}
\caption{Average monogamy score, $\delta_{\mathcal{B}}^{\text{avg}}$  (vertical axis) against  negativity of the initial pure state, \(|\psi\rangle\), ${N}$ (horizontal axis). All other specifications are same as in Fig. \ref{fig:initial_ent_case}. Both the  axes are dimensionless.}  
\label{fig:avg_nonmax}
\end{figure}

$3.$ Like the case of post-selection, $\delta^{\text{avg}}_{\mathcal{B}}$ also gets maximized for the moderate value of $\lambda$ and the value itself increases with the increase of entanglement in the input state as shown in Fig. \ref{fig:avg_nonmax}. However, for low values of $\lambda$ nonmaximally entangled states again turn out to be optimal in terms of monogamy score. 

\begin{figure}
\includegraphics [width=\linewidth]{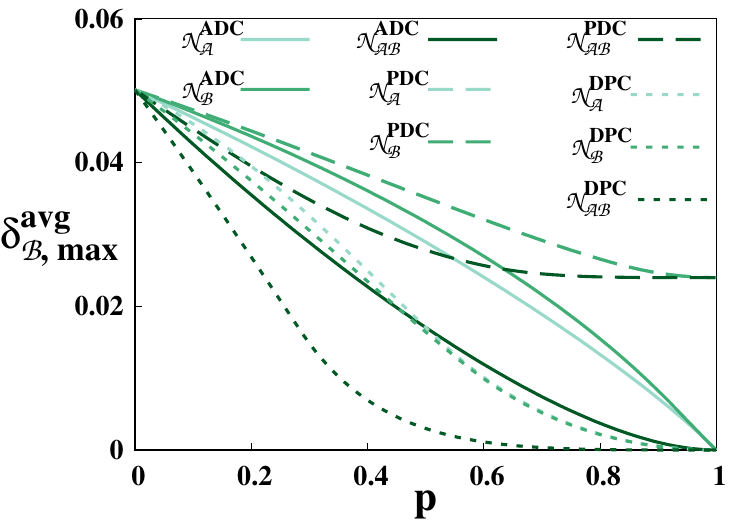}
\caption{Average maximized monogamy  score, $\delta_{\mathcal{B},{\text{max}}}^{\text{avg}}$ (ordinate) against local noise parameter, $p$ (abscissa). Here the maximization is performed over  unsharp parameter, $\lambda$ as well as the parameters in the auxiliary qubits.  The initial resource state is the nonmaximally entangled state,  $\ket {\psi}$ with $z=\frac{\pi}{8}$. Same specifications of linetypes and colors are followed for $\mathcal{N}_{\mathcal{A}}$, $\mathcal{N}_{\mathcal{B}}$, $\mathcal{N}_{\mathcal{A}\mathcal{B}}$ and ADC, PDC, DPC as in  Fig. \ref{fig:avg_mono_lam}. Both the ordinate and abscissa are dimensionless.}
\label{fig:avg_3party}
\end{figure}

$4.$ Among these three prototypical  channels, ADC and PDC have less severe influence on the shareability of multipartite entanglement in comparison to the DPC. Specifically, we observe that if we fix the position of noise in the initial state, the monogamy scores can provide a hierarchy among different noisy channels, given by
\begin{eqnarray}
    \delta^{\text{avg}}_{\mathcal{B}} \big(\mathcal{N}^{\text{PDC}} (\rho_{\mathcal{A}\mathcal{B}\{{B}_1\}})\big) &>& \delta^{\text{avg}}_{\mathcal{B}} \big(\mathcal{N}^{\text{ADC}} (\rho_{\mathcal{A}\mathcal{B}\{{B}_1\}})\big)\nonumber\\ &>& \delta^{\text{avg}}_{\mathcal{B}} \big(\mathcal{N}^{\text{DPC}} (\rho_{\mathcal{A}\mathcal{B}\{{B}_1\}})\big).
\end{eqnarray}
It turns out to be independent of the shared entanglement in inputs. To eliminate the dependence of $\lambda$, let us now optimize $\delta^{\text{avg}}_{\mathcal{B}}$  with respect to $\lambda$ along with $\theta$ and $\phi$ of the auxiliary qubit which we refer to $\delta^{\text{avg}}_{\mathcal{B},\max}$,  dependent only on $p$ and the initial entanglement. In this situation, as the strength of the noise increases, $\delta^{\text{avg}}_{\mathcal{B},\max}(\rho_{\mathcal{A}\mathcal{B}\{{B}_1\}}, p)$ decreases monotonically for different kinds of noisy channels (see Fig. \ref{fig:avg_3party}).
The hierarchies mentioned for post-selection holds even when we optimize  $\lambda$. Another interesting observation include the change of curvatures in $\delta^{\text{avg}}_{\mathcal{B}}(\rho_{\mathcal{A}\mathcal{B}\{{B}_1\}}, p, \lambda)$ with $p$ when a single channel acts on the state and when both the channels are noisy except the DPC. A possible explanation can be due to the fact that $\delta^{\text{avg}}_{\mathcal{B}}(\rho_{\mathcal{A}\mathcal{B}\{{B}_1\}}, p, \lambda)$ in the former case depends on $p$ while  $\delta^{\text{avg}}_{\mathcal{B}}$ depends on $p^2$ in the latter case, resulting to a contrasting behavior. 

\begin{figure}
\includegraphics [width=\linewidth]{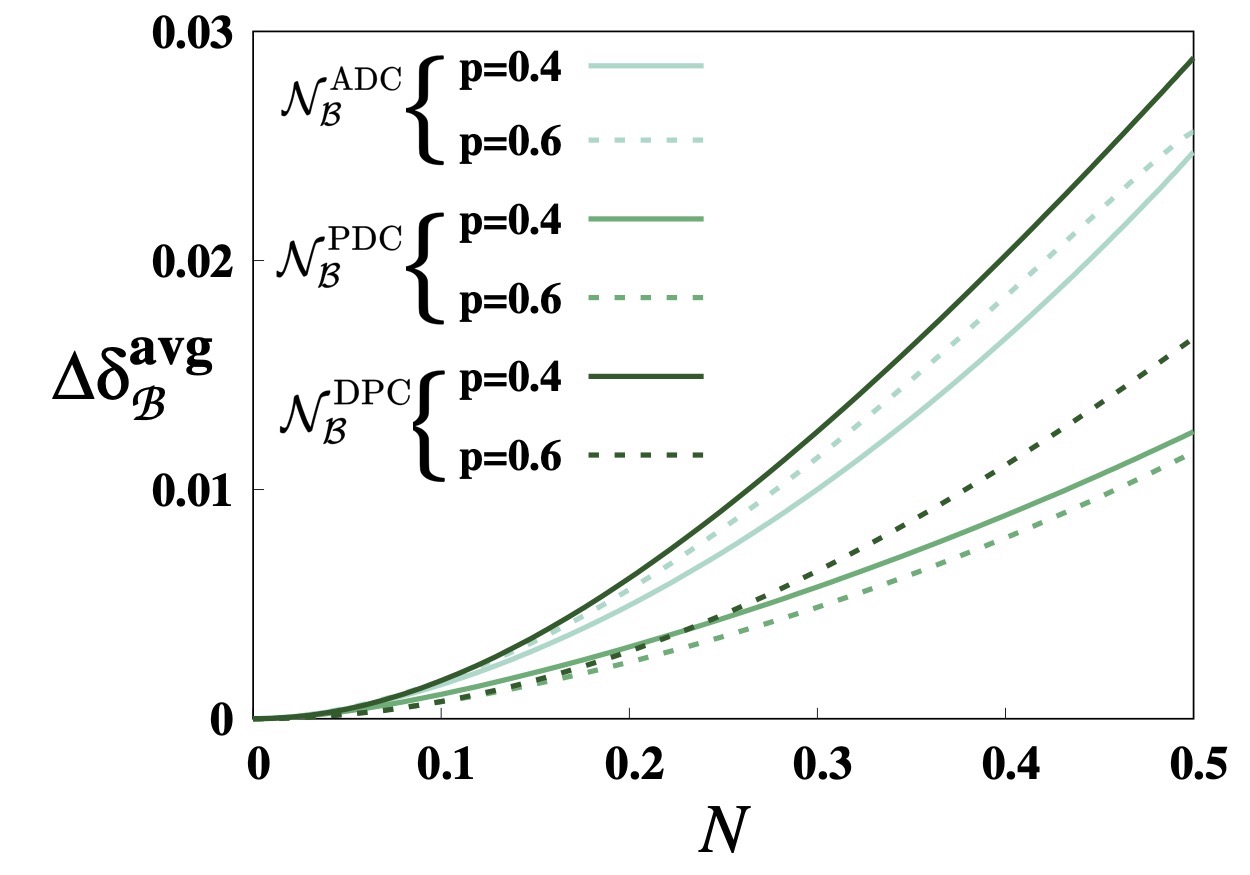}
\caption{$\Delta\delta^{\text{avg}}_{\mathcal{B}}$ (given in Eq. (\ref{eq:robust_equ})) (vertical axis) vs negativity, ${N}$ (horizontal axis) of initial resource state, $\ket\psi$. Qubits are effected by ADC, PDC, DPC (sequentially light green to dark green) 
in the form of local noise, $p=0.4$ (solid) and $p=0.6$ (dotted) in the unidirectional case. Both the axes are dimensionless.} 
\label{fig:U_robust_mono}
\end{figure}

\textbf{Robustness of monogamy against noise.} To assess quantitatively the resilience of the monogamy score when transitioning from the situation with noise on a single party to both the parties, we introduce a quantity, 
\begin{eqnarray}
\Delta\delta_{\mathcal{B}}^\text{avg}
&=&\delta_{\mathcal{B},\text{max}}^{\text{avg}}(\rho_{\mathcal{A}\mathcal{B}\{\boldsymbol{B}\}},p)|_{\mathbf{U}(\mathcal{B})_n} \nonumber \\ &-&\delta_{\mathcal{B},\text{max}}^{\text{avg}}(\rho_{\mathcal{A}\mathcal{B}\{\boldsymbol{B}\}},p)|_{\mathbf{U}(\mathcal{A}\mathcal{B})_n},
\label{eq:robust_equ}
\end{eqnarray}
whose positive value guarantees that the propagation of entanglement is less disturbed by a single noisy channel than that of the double noisy channels. As observed before, $\Delta\delta_{\mathcal{B}}^\text{avg}$ monotonically increases with the entanglement of the initial state irrespective of the channel acted on the initial state. Further, with the increase of $p$, $\Delta\delta_{\mathcal{B}}^\text{avg}$ decreases which implies that the decrease of $\delta^{\text{avg}}_{\mathcal{B}}$ in presence of a single noisy channel is much more drastic than that of double noisy channels although both the quantities involved decreases in presence of noise. Again, the phase damping channel has the least destructive impact on entanglement distribution compared to the other channels. Interestingly, it demonstrates that  more entanglement possessed by the initial state, the influence of noise, especially in presence of both the channels being noisy, seems to be more. Precisely, from the data used in Fig. \ref{fig:U_robust_mono}, we find that when the initial entanglement is maximum (i.e., $N(\rho_{\mathcal{A}\mathcal{B}}) = 0.5 $), the  depolarizing channels acted on both the nodes with \(p=0.4\) affect $4.38 \%$ more compared to the single channel while in case of the initial state having entanglement content $0.2$, the percentage reduces to $2.9 \%$, thereby illustrating less destructive effect in case of a weakly entangled state.
\section{Sharing entanglement: Bidirectional vs Unidirectional}
\label{sec:uni_vs_bi}
To make the entanglement propagation more symmetric, we now add auxiliary systems, both in $\mathcal{A}$ and $\mathcal{B}$ parties, who perform unsharp measurements. For simplicity, the unsharpness parameters at both the nodes are considered to be same. We study the monogamy score of the four party resulting state denoted as, $\delta_{\mathcal{B}}\big(\rho^{\{3\}\{3\}}_{\mathcal{A}\{A_{1}\}\mathcal{B}\{B_{1}\}},p,\lambda\big)$, when both the parties obtain the outcome $M^{3}$ in presence of noise affecting either party $\mathcal{A}$ or $\mathcal{B}$ or both. Let us enumerate some critical observations that either deviate from the unidirectional case or remain consistent with the unidirectional scenario.
\begin{figure}
\includegraphics [width=\linewidth]{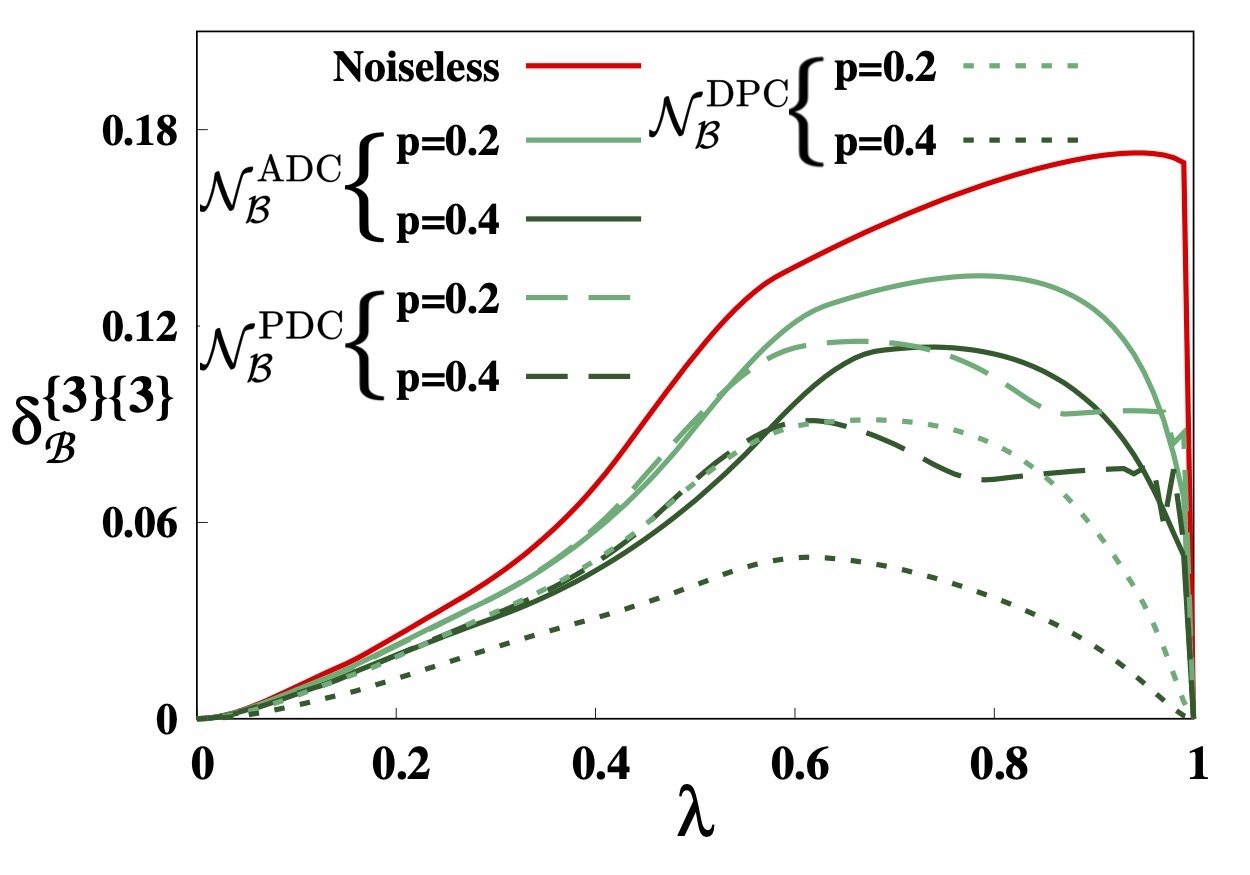}
\caption{Monogamy score, $\delta_{\mathcal{B}}^{\{3\}\{3\}}$ (ordinate)   of the four-partite entangled state generated from  the bidirectional protocol $\mathbf{B}(\mathcal{B})_1^1$ against unsharpness strength $\lambda$ (abscissa). The outcomes of both  the measurements performed by \(\mathcal{A}\) and \(\mathcal{B}\)  are chosen to be $M^3$. Different curves correspond to the different values of the noise content and noisy channels. All other specifications, including color coding and line types, are same as in Fig. \ref{fig:u1_case}. Both the axes are dimensionless.}   
\label{fig:selective_bidirec_case}
\end{figure}

$1.$ Noise again leads to a suppression in entanglement-distribution in selective bidirectional scenario. We observe that when the noise acts on one of nodes of $\ket{\psi}$, say $\mathcal{B}$, the corresponding input state can produce a lower monogamy score in comparison to the noiseless scenario for all values of $\lambda$ and for all kinds of prototypical noise models considered in this work. The maximum monogamy score is again achieved for moderate values of $\lambda$ with a fixed amount of entanglement
as shown in Fig. \ref{fig:selective_bidirec_case}. It can possibly be argued that such a behavior occurs due to the dual influence of single noisy channel affecting node $\mathcal{B}$ and unsharp measurement in both $\mathcal{A}$ and $\mathcal{B}$.

$2.$ Like unidirectional case, nonmaximally entangled state performs better as the initial resource than that of the the maximally entangled one. For example, when ADC acts on $\mathcal{B}$ with $p=0.2$ and unsharpness value $\lambda=0.8$, we find that $\delta_{\mathcal{B}}^{\{3\}\{3\}}$ reaches to $0.161$ when the initial state possess entanglement $N=0.176$ while it becomes $0.137$ for maximally entangled state. It establishes that in presence of noise, there always exists a fixed entangled state which acquires maximum $\delta_{\mathcal{B}}^{\{3\}\{3\}}$ for a given unsharp parameter which may not be the maximally entangled state. Like the unidirectional case, when both high and low unsharpness are present in measurements, maximal entanglement shareability power of the four-party state in networks can be accomplished via nonmaximally entangled states which is not the case for moderate $\lambda$ values.
\begin{figure}
\includegraphics [width=\linewidth]{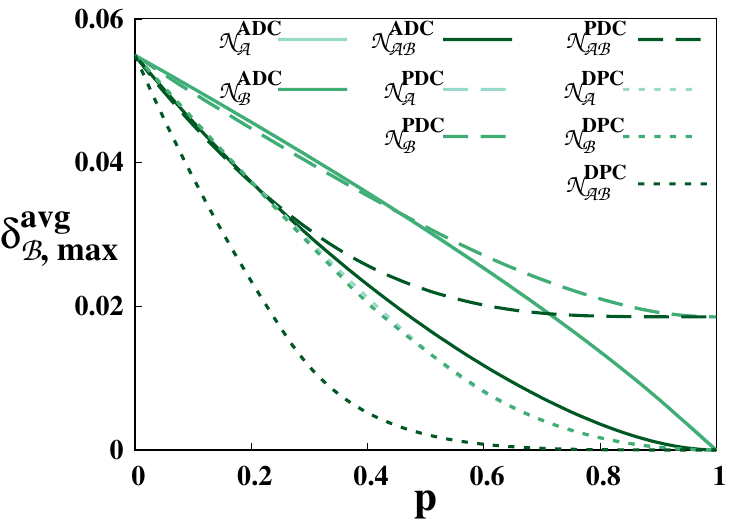}
\caption{Maximized average monogamy score, $\delta_{\mathcal{B},\max}^{\text{avg}}$ (vertical axis) vs $p$ (horizontal axis) in the bidirectional protocol $\mathbf{B}_1^1$ for local noises in $\mathcal{A(B)}$ or in both $\mathcal{A}$ and $\mathcal{B}$. All the  other specifications are same as in Fig. \ref{fig:avg_3party}. Both the axes are dimensionless.}  
\label{fig:avg_bidirec_p}
\end{figure}

$3.$ The comparisons between different channels are carried out by  averaging over the measurement outcomes which is in this case sixteen and by optimizing over $\{\theta_{i},\phi_{i}\}_{i=1}^{2}$ of the auxiliary qubits and unsharpness $\lambda$ in measurements. The shareability quantifier,
$\delta_{\mathcal{B},\max}^{\text{avg}}\big(\rho_{\mathcal{A}\{A_{1}\}\mathcal{B}\{B_{1}\}},p\big)$ decreases with the increase of $p$ and their curvature again changes if a single node of $\ket{\psi}$ or both the nodes of $\ket{\psi}$ are affected with noise except DPC. The impact of DPC on distributing entanglement is worst for the entanglement propagation than other two paradigmatic channels. By comparing unidirectional and bidirectional cases (by comparing Fig. \ref{fig:avg_3party} and \ref{fig:avg_bidirec_p}), we observe two contrasting features -- \textbf{a.} Due to symmetry, in the bidirectional case, entanglement monogamy scores are always equal when the channel acts on $\mathcal{A}$ or $\mathcal{B}$. \textbf{b.} Let us compare a scenario when ADC or PDC acts on one of the nodes, for weak noise strength. We observe that the monogamy score obtained in the ADC case takes higher value compared to the PDC case when the initial shared state possess a moderate amount of entanglement. The opposite picture emerges for the unidirectional case which is true for all values of $p$ and the initial entanglement. 
\begin{figure}
\includegraphics [width=\linewidth]{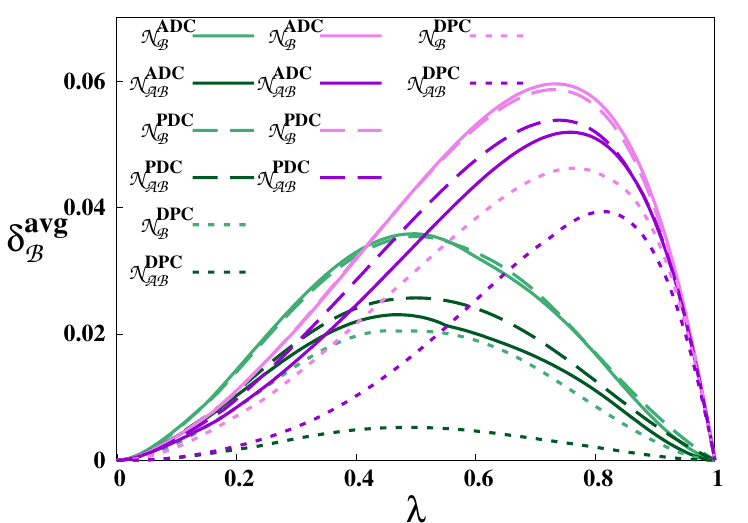}
\caption{Average  maximized monogamy score, $\delta^\text{avg}_\mathcal{B}$ ($y$-axis) against $\lambda$ ($x$-axis) for fixed $p=0.4$. The output state is four-partite  entangled state generated  from both the bidirectional, $\mathbf{B}_1^1$ (green shades) and unidirectional $\mathbf{U}_2$ (violet shades) propagation protocols. Local ADC (solid), PDC (dashed), DPC (dotted) act  only on $\mathcal{B}$ (bright) or on both $\mathcal{A}$  and \(\mathcal{B}\) (dark) nodes. Initial resource state is $\ket\psi$ for $z=\frac{\pi}{8}$. Clearly, the unidirectional protocols are more advantageous for sharing entanglement in multiple nodes than the bidirectional ones. Both the axes are dimensionless.}  
\label{fig:4party_uni_diff_lam}
\end{figure}

$4.$ We endeavour to conduct a comparative analysis of the characteristics shown by four-party genuine entangled states created from both the unidirectional and the bidirectional pictures. In the unidirectional case, two unsharp measurements are performed sequentially by $\mathcal{B}$ while in the bidirectional situation, both $\mathcal{A}$ and $\mathcal{B}$ perform unsharp measurement. In the former case, if noise acts on, say party $\mathcal{B}$, there is a possibility that with the increase of number of measurements by $\mathcal{B}$, entanglement shareability in networks gets less affected than the bidirectional protocol. We find that this is indeed the case. The higher average monogamy score, indicating higher entanglement propagation, is observed when a single party performs all the measurements in presence of noise and by suitably tuning unsharpness parameter in the measurement compared to the bidirectional situation (see Fig. \ref{fig:4party_uni_diff_lam}). This result is independent of the noise model affecting the resource state. As discussed in the preceding sections, the effect of depolarizing noise on the entanglement of the four party unidirectional scheme is most pronounced compared to other channels like PDC and ADC.

\section{Conclusion}
\label{sec:conc}

Distributing both classical and quantum information among multiple nodes is essential in constructing quantum communication networks which can also be an integral part of quantum circuits.  The distribution of entanglement among nodes is the fundamental step necessary for the establishment of such networks. 
Several methods are developed to distribute and certify entanglement across several sites, despite the fact that the majority of earlier research either deal with initial pure states or mixed states that are eventually purified when a significant number of copies are accessible.

We proposed a set-up for propagating entanglement in networks with an arbitrary number of sites and  geometry  in which the channels connecting the source and nodes are noisy and the measurement apparatus are also imperfect. Our goal was to demonstrate how to increase the shareability of entanglement in networks in presence of dual defects. The shareability power of the output state is measured by its monogamy score.
We presented a recursion relation for creating a multipartite state after arbitrary number of rounds starting from bipartite noisy state, suitably prepared auxiliary single qubit states and properly tuned unsharp measurements. We discovered that after the critical value of the unsharp parameter present in the measurement, a noisy resource state can achieve a higher monogamy score than that can obtained in a noiseless scenario, provided the outcome of the measurement is post-selected and one of the local channel connecting the node is amplitude damping and that nodes only perform measurement sequentially. Further, we also demonstrated that a nonmaximally entangled initial state have a higher monogamy score than that of the maximally entangled state for a fixed unsharpness in measurements.  When compared to dephasing and amplitude damping channels, depolarizing channels had the most detrimental effects on entanglement shareability of all archetypal noisy channels analysed in this study. 
When noisy channels act on both parties rather than a single site, the shareability capacity of entanglement in the resulting state always decreases, although this destructive effect of noisy initially entangled state can be eliminated by adding more auxiliary qubits and performing more unsharp measurements.

\acknowledgements
 
We acknowledge the support from Interdisciplinary Cyber Physical Systems (ICPS) program of the Department of Science and Technology (DST), India, Grant No.: DST/ICPS/QuST/Theme- 1/2019/23. We  acknowledge the use of \href{https://github.com/titaschanda/QIClib}{QIClib} -- a modern C++ library for general purpose quantum information processing and quantum computing (\url{https://titaschanda.github.io/QIClib}) and cluster computing facility at Harish-Chandra Research Institute.

\bibliographystyle{apsrev4-1}
	\bibliography{bib.bib}

\begin{thebibliography}{96}%
\makeatletter
\providecommand \@ifxundefined [1]{%
 \@ifx{#1\undefined}
}%
\providecommand \@ifnum [1]{%
 \ifnum #1\expandafter \@firstoftwo
 \else \expandafter \@secondoftwo
 \fi
}%
\providecommand \@ifx [1]{%
 \ifx #1\expandafter \@firstoftwo
 \else \expandafter \@secondoftwo
 \fi
}%
\providecommand \natexlab [1]{#1}%
\providecommand \enquote  [1]{``#1''}%
\providecommand \bibnamefont  [1]{#1}%
\providecommand \bibfnamefont [1]{#1}%
\providecommand \citenamefont [1]{#1}%
\providecommand \href@noop [0]{\@secondoftwo}%
\providecommand \href [0]{\begingroup \@sanitize@url \@href}%
\providecommand \@href[1]{\@@startlink{#1}\@@href}%
\providecommand \@@href[1]{\endgroup#1\@@endlink}%
\providecommand \@sanitize@url [0]{\catcode `\\12\catcode `\$12\catcode
  `\&12\catcode `\#12\catcode `\^12\catcode `\_12\catcode `\%12\relax}%
\providecommand \@@startlink[1]{}%
\providecommand \@@endlink[0]{}%
\providecommand \url  [0]{\begingroup\@sanitize@url \@url }%
\providecommand \@url [1]{\endgroup\@href {#1}{\urlprefix }}%
\providecommand \urlprefix  [0]{URL }%
\providecommand \Eprint [0]{\href }%
\providecommand \doibase [0]{http://dx.doi.org/}%
\providecommand \selectlanguage [0]{\@gobble}%
\providecommand \bibinfo  [0]{\@secondoftwo}%
\providecommand \bibfield  [0]{\@secondoftwo}%
\providecommand \translation [1]{[#1]}%
\providecommand \BibitemOpen [0]{}%
\providecommand \bibitemStop [0]{}%
\providecommand \bibitemNoStop [0]{.\EOS\space}%
\providecommand \EOS [0]{\spacefactor3000\relax}%
\providecommand \BibitemShut  [1]{\csname bibitem#1\endcsname}%
\let\auto@bib@innerbib\@empty
\bibitem [{\citenamefont {Horodecki}\ \emph {et~al.}(2009)\citenamefont
  {Horodecki}, \citenamefont {Horodecki}, \citenamefont {Horodecki},\ and\
  \citenamefont {Horodecki}}]{horo_09_ent}%
  \BibitemOpen
  \bibfield  {author} {\bibinfo {author} {\bibfnamefont {R.}~\bibnamefont
  {Horodecki}}, \bibinfo {author} {\bibfnamefont {P.}~\bibnamefont
  {Horodecki}}, \bibinfo {author} {\bibfnamefont {M.}~\bibnamefont
  {Horodecki}}, \ and\ \bibinfo {author} {\bibfnamefont {K.}~\bibnamefont
  {Horodecki}},\ }\href {\doibase 10.1103/RevModPhys.81.865} {\bibfield
  {journal} {\bibinfo  {journal} {Rev. Mod. Phys.}\ }\textbf {\bibinfo {volume}
  {81}},\ \bibinfo {pages} {865} (\bibinfo {year} {2009})}\BibitemShut
  {NoStop}%
\bibitem [{\citenamefont {Ollivier}\ and\ \citenamefont
  {Zurek}(2001)}]{zurek01discord}%
  \BibitemOpen
  \bibfield  {author} {\bibinfo {author} {\bibfnamefont {H.}~\bibnamefont
  {Ollivier}}\ and\ \bibinfo {author} {\bibfnamefont {W.~H.}\ \bibnamefont
  {Zurek}},\ }\href {\doibase 10.1103/PhysRevLett.88.017901} {\bibfield
  {journal} {\bibinfo  {journal} {Phys. Rev. Lett.}\ }\textbf {\bibinfo
  {volume} {88}},\ \bibinfo {pages} {017901} (\bibinfo {year}
  {2001})}\BibitemShut {NoStop}%
\bibitem [{\citenamefont {Bera}\ \emph {et~al.}(2018)\citenamefont {Bera},
  \citenamefont {Das}, \citenamefont {Sadhukhan}, \citenamefont {Roy},
  \citenamefont {Sen(De)},\ and\ \citenamefont {Sen}}]{Bera2018discord}%
  \BibitemOpen
  \bibfield  {author} {\bibinfo {author} {\bibfnamefont {A.}~\bibnamefont
  {Bera}}, \bibinfo {author} {\bibfnamefont {T.}~\bibnamefont {Das}}, \bibinfo
  {author} {\bibfnamefont {D.}~\bibnamefont {Sadhukhan}}, \bibinfo {author}
  {\bibfnamefont {S.~S.}\ \bibnamefont {Roy}}, \bibinfo {author} {\bibfnamefont
  {A.}~\bibnamefont {Sen(De)}}, \ and\ \bibinfo {author} {\bibfnamefont
  {U.}~\bibnamefont {Sen}},\ }\href {\doibase 10.1088/1361-6633/aa872f}
  {\bibfield  {journal} {\bibinfo  {journal} {Reports on Progress in Physics}\
  }\textbf {\bibinfo {volume} {81}},\ \bibinfo {pages} {024001} (\bibinfo
  {year} {2018})}\BibitemShut {NoStop}%
\bibitem [{\citenamefont {Streltsov}\ \emph {et~al.}(2017)\citenamefont
  {Streltsov}, \citenamefont {Adesso},\ and\ \citenamefont
  {Plenio}}]{stresltsov_17_coherence}%
  \BibitemOpen
  \bibfield  {author} {\bibinfo {author} {\bibfnamefont {A.}~\bibnamefont
  {Streltsov}}, \bibinfo {author} {\bibfnamefont {G.}~\bibnamefont {Adesso}}, \
  and\ \bibinfo {author} {\bibfnamefont {M.~B.}\ \bibnamefont {Plenio}},\
  }\href {\doibase 10.1103/RevModPhys.89.041003} {\bibfield  {journal}
  {\bibinfo  {journal} {Rev. Mod. Phys.}\ }\textbf {\bibinfo {volume} {89}},\
  \bibinfo {pages} {041003} (\bibinfo {year} {2017})}\BibitemShut {NoStop}%
\bibitem [{\citenamefont {Bennett}\ \emph {et~al.}(1993)\citenamefont
  {Bennett}, \citenamefont {Brassard}, \citenamefont {Cr\'epeau}, \citenamefont
  {Jozsa}, \citenamefont {Peres},\ and\ \citenamefont
  {Wootters}}]{Bennett_1993}%
  \BibitemOpen
  \bibfield  {author} {\bibinfo {author} {\bibfnamefont {C.~H.}\ \bibnamefont
  {Bennett}}, \bibinfo {author} {\bibfnamefont {G.}~\bibnamefont {Brassard}},
  \bibinfo {author} {\bibfnamefont {C.}~\bibnamefont {Cr\'epeau}}, \bibinfo
  {author} {\bibfnamefont {R.}~\bibnamefont {Jozsa}}, \bibinfo {author}
  {\bibfnamefont {A.}~\bibnamefont {Peres}}, \ and\ \bibinfo {author}
  {\bibfnamefont {W.~K.}\ \bibnamefont {Wootters}},\ }\href {\doibase
  10.1103/PhysRevLett.70.1895} {\bibfield  {journal} {\bibinfo  {journal}
  {Phys. Rev. Lett.}\ }\textbf {\bibinfo {volume} {70}},\ \bibinfo {pages}
  {1895} (\bibinfo {year} {1993})}\BibitemShut {NoStop}%
\bibitem [{\citenamefont {Bouwmeester}\ \emph {et~al.}(1997)\citenamefont
  {Bouwmeester}, \citenamefont {Pan}, \citenamefont {Mattle}, \citenamefont
  {Eibl}, \citenamefont {Weinfurter},\ and\ \citenamefont
  {Zeilinger}}]{Bouwmeester_1997}%
  \BibitemOpen
  \bibfield  {author} {\bibinfo {author} {\bibfnamefont {D.}~\bibnamefont
  {Bouwmeester}}, \bibinfo {author} {\bibfnamefont {J.-W.}\ \bibnamefont
  {Pan}}, \bibinfo {author} {\bibfnamefont {K.}~\bibnamefont {Mattle}},
  \bibinfo {author} {\bibfnamefont {M.}~\bibnamefont {Eibl}}, \bibinfo {author}
  {\bibfnamefont {H.}~\bibnamefont {Weinfurter}}, \ and\ \bibinfo {author}
  {\bibfnamefont {A.}~\bibnamefont {Zeilinger}},\ }\href {\doibase
  10.1038/37539} {\bibfield  {journal} {\bibinfo  {journal} {Nature}\ }\textbf
  {\bibinfo {volume} {390}},\ \bibinfo {pages} {575} (\bibinfo {year}
  {1997})}\BibitemShut {NoStop}%
\bibitem [{\citenamefont {Murao}\ \emph {et~al.}(1999)\citenamefont {Murao},
  \citenamefont {Jonathan}, \citenamefont {Plenio},\ and\ \citenamefont
  {Vedral}}]{Murao_1999}%
  \BibitemOpen
  \bibfield  {author} {\bibinfo {author} {\bibfnamefont {M.}~\bibnamefont
  {Murao}}, \bibinfo {author} {\bibfnamefont {D.}~\bibnamefont {Jonathan}},
  \bibinfo {author} {\bibfnamefont {M.~B.}\ \bibnamefont {Plenio}}, \ and\
  \bibinfo {author} {\bibfnamefont {V.}~\bibnamefont {Vedral}},\ }\href
  {\doibase 10.1103/PhysRevA.59.156} {\bibfield  {journal} {\bibinfo  {journal}
  {Phys. Rev. A}\ }\textbf {\bibinfo {volume} {59}},\ \bibinfo {pages} {156}
  (\bibinfo {year} {1999})}\BibitemShut {NoStop}%
\bibitem [{\citenamefont {Grudka}(2004)}]{grudka_2004}%
  \BibitemOpen
  \bibfield  {author} {\bibinfo {author} {\bibfnamefont {A.}~\bibnamefont
  {Grudka}},\ }\href@noop {} {\bibfield  {journal} {\bibinfo  {journal} {Acta
  Physica Slovaca}\ }\textbf {\bibinfo {volume} {54}} (\bibinfo {year}
  {2004})}\BibitemShut {NoStop}%
\bibitem [{\citenamefont {Sen(De)}\ and\ \citenamefont
  {Sen}(2010{\natexlab{a}})}]{Sen(de)_2010}%
  \BibitemOpen
  \bibfield  {author} {\bibinfo {author} {\bibfnamefont {A.}~\bibnamefont
  {Sen(De)}}\ and\ \bibinfo {author} {\bibfnamefont {U.}~\bibnamefont {Sen}},\
  }\href {\doibase 10.1103/PhysRevA.81.012308} {\bibfield  {journal} {\bibinfo
  {journal} {Phys. Rev. A}\ }\textbf {\bibinfo {volume} {81}},\ \bibinfo
  {pages} {012308} (\bibinfo {year} {2010}{\natexlab{a}})}\BibitemShut
  {NoStop}%
\bibitem [{\citenamefont {Bennett}\ \emph {et~al.}(2005)\citenamefont
  {Bennett}, \citenamefont {Hayden}, \citenamefont {Leung}, \citenamefont
  {Shor},\ and\ \citenamefont {Winter}}]{Bennett_2005}%
  \BibitemOpen
  \bibfield  {author} {\bibinfo {author} {\bibfnamefont {C.}~\bibnamefont
  {Bennett}}, \bibinfo {author} {\bibfnamefont {P.}~\bibnamefont {Hayden}},
  \bibinfo {author} {\bibfnamefont {D.}~\bibnamefont {Leung}}, \bibinfo
  {author} {\bibfnamefont {P.}~\bibnamefont {Shor}}, \ and\ \bibinfo {author}
  {\bibfnamefont {A.}~\bibnamefont {Winter}},\ }\href {\doibase
  10.1109/TIT.2004.839476} {\bibfield  {journal} {\bibinfo  {journal} {IEEE
  Transactions on Information Theory}\ }\textbf {\bibinfo {volume} {51}},\
  \bibinfo {pages} {56} (\bibinfo {year} {2005})}\BibitemShut {NoStop}%
\bibitem [{\citenamefont {Pati}(2000)}]{pati_2000}%
  \BibitemOpen
  \bibfield  {author} {\bibinfo {author} {\bibfnamefont {A.~K.}\ \bibnamefont
  {Pati}},\ }\href {\doibase 10.1103/PhysRevA.63.014302} {\bibfield  {journal}
  {\bibinfo  {journal} {Phys. Rev. A}\ }\textbf {\bibinfo {volume} {63}},\
  \bibinfo {pages} {014302} (\bibinfo {year} {2000})}\BibitemShut {NoStop}%
\bibitem [{\citenamefont {Bennett}\ and\ \citenamefont
  {Wiesner}(1992)}]{Bennett_1992}%
  \BibitemOpen
  \bibfield  {author} {\bibinfo {author} {\bibfnamefont {C.~H.}\ \bibnamefont
  {Bennett}}\ and\ \bibinfo {author} {\bibfnamefont {S.~J.}\ \bibnamefont
  {Wiesner}},\ }\href {\doibase 10.1103/PhysRevLett.69.2881} {\bibfield
  {journal} {\bibinfo  {journal} {Phys. Rev. Lett.}\ }\textbf {\bibinfo
  {volume} {69}},\ \bibinfo {pages} {2881} (\bibinfo {year}
  {1992})}\BibitemShut {NoStop}%
\bibitem [{\citenamefont {Mattle}\ \emph {et~al.}(1996)\citenamefont {Mattle},
  \citenamefont {Weinfurter}, \citenamefont {Kwiat},\ and\ \citenamefont
  {Zeilinger}}]{Mattle_1996}%
  \BibitemOpen
  \bibfield  {author} {\bibinfo {author} {\bibfnamefont {K.}~\bibnamefont
  {Mattle}}, \bibinfo {author} {\bibfnamefont {H.}~\bibnamefont {Weinfurter}},
  \bibinfo {author} {\bibfnamefont {P.~G.}\ \bibnamefont {Kwiat}}, \ and\
  \bibinfo {author} {\bibfnamefont {A.}~\bibnamefont {Zeilinger}},\ }\href
  {\doibase 10.1103/PhysRevLett.76.4656} {\bibfield  {journal} {\bibinfo
  {journal} {Phys. Rev. Lett.}\ }\textbf {\bibinfo {volume} {76}},\ \bibinfo
  {pages} {4656} (\bibinfo {year} {1996})}\BibitemShut {NoStop}%
\bibitem [{\citenamefont {Bru\ss{}}\ \emph {et~al.}(2004)\citenamefont
  {Bru\ss{}}, \citenamefont {D'Ariano}, \citenamefont {Lewenstein},
  \citenamefont {Macchiavello}, \citenamefont {Sen(De)},\ and\ \citenamefont
  {Sen}}]{Bruss_2004}%
  \BibitemOpen
  \bibfield  {author} {\bibinfo {author} {\bibfnamefont {D.}~\bibnamefont
  {Bru\ss{}}}, \bibinfo {author} {\bibfnamefont {G.~M.}\ \bibnamefont
  {D'Ariano}}, \bibinfo {author} {\bibfnamefont {M.}~\bibnamefont
  {Lewenstein}}, \bibinfo {author} {\bibfnamefont {C.}~\bibnamefont
  {Macchiavello}}, \bibinfo {author} {\bibfnamefont {A.}~\bibnamefont
  {Sen(De)}}, \ and\ \bibinfo {author} {\bibfnamefont {U.}~\bibnamefont
  {Sen}},\ }\href {\doibase 10.1103/PhysRevLett.93.210501} {\bibfield
  {journal} {\bibinfo  {journal} {Phys. Rev. Lett.}\ }\textbf {\bibinfo
  {volume} {93}},\ \bibinfo {pages} {210501} (\bibinfo {year}
  {2004})}\BibitemShut {NoStop}%
\bibitem [{\citenamefont {Bru{\ss}}\ \emph {et~al.}(2006)\citenamefont
  {Bru{\ss}}, \citenamefont {Lewenstein}, \citenamefont {Sen(De)},
  \citenamefont {Sen}, \citenamefont {D{\textquotesingle}ariano},\ and\
  \citenamefont {Macchiavello}}]{BRU__2006}%
  \BibitemOpen
  \bibfield  {author} {\bibinfo {author} {\bibfnamefont {D.}~\bibnamefont
  {Bru{\ss}}}, \bibinfo {author} {\bibfnamefont {M.}~\bibnamefont
  {Lewenstein}}, \bibinfo {author} {\bibfnamefont {A.}~\bibnamefont {Sen(De)}},
  \bibinfo {author} {\bibfnamefont {U.}~\bibnamefont {Sen}}, \bibinfo {author}
  {\bibfnamefont {G.~M.}\ \bibnamefont {D{\textquotesingle}ariano}}, \ and\
  \bibinfo {author} {\bibfnamefont {C.}~\bibnamefont {Macchiavello}},\ }\href
  {\doibase 10.1142/s0219749906001888} {\bibfield  {journal} {\bibinfo
  {journal} {International Journal of Quantum Information}\ }\textbf {\bibinfo
  {volume} {04}},\ \bibinfo {pages} {415} (\bibinfo {year} {2006})}\BibitemShut
  {NoStop}%
\bibitem [{\citenamefont {De}\ and\ \citenamefont {Sen}(2011)}]{de_2011}%
  \BibitemOpen
  \bibfield  {author} {\bibinfo {author} {\bibfnamefont {A.~S.}\ \bibnamefont
  {De}}\ and\ \bibinfo {author} {\bibfnamefont {U.}~\bibnamefont {Sen}},\
  }\href@noop {} {\enquote {\bibinfo {title} {Quantum advantage in
  communication networks},}\ } (\bibinfo {year} {2011}),\ \Eprint
  {http://arxiv.org/abs/1105.2412} {arXiv:1105.2412 [quant-ph]} \BibitemShut
  {NoStop}%
\bibitem [{\citenamefont {Horodecki}\ and\ \citenamefont
  {Piani}(2012)}]{Horodecki_2012}%
  \BibitemOpen
  \bibfield  {author} {\bibinfo {author} {\bibfnamefont {M.}~\bibnamefont
  {Horodecki}}\ and\ \bibinfo {author} {\bibfnamefont {M.}~\bibnamefont
  {Piani}},\ }\href {\doibase 10.1088/1751-8113/45/10/105306} {\bibfield
  {journal} {\bibinfo  {journal} {Journal of Physics A: Mathematical and
  Theoretical}\ }\textbf {\bibinfo {volume} {45}},\ \bibinfo {pages} {105306}
  (\bibinfo {year} {2012})}\BibitemShut {NoStop}%
\bibitem [{\citenamefont {Shadman}\ \emph {et~al.}(2012)\citenamefont
  {Shadman}, \citenamefont {Kampermann}, \citenamefont {Bru\ss{}},\ and\
  \citenamefont {Macchiavello}}]{Shadman_2012}%
  \BibitemOpen
  \bibfield  {author} {\bibinfo {author} {\bibfnamefont {Z.}~\bibnamefont
  {Shadman}}, \bibinfo {author} {\bibfnamefont {H.}~\bibnamefont {Kampermann}},
  \bibinfo {author} {\bibfnamefont {D.}~\bibnamefont {Bru\ss{}}}, \ and\
  \bibinfo {author} {\bibfnamefont {C.}~\bibnamefont {Macchiavello}},\ }\href
  {\doibase 10.1103/PhysRevA.85.052306} {\bibfield  {journal} {\bibinfo
  {journal} {Phys. Rev. A}\ }\textbf {\bibinfo {volume} {85}},\ \bibinfo
  {pages} {052306} (\bibinfo {year} {2012})}\BibitemShut {NoStop}%
\bibitem [{\citenamefont {Das}\ \emph {et~al.}(2014)\citenamefont {Das},
  \citenamefont {Prabhu}, \citenamefont {Sen(De)},\ and\ \citenamefont
  {Sen}}]{Das_2014}%
  \BibitemOpen
  \bibfield  {author} {\bibinfo {author} {\bibfnamefont {T.}~\bibnamefont
  {Das}}, \bibinfo {author} {\bibfnamefont {R.}~\bibnamefont {Prabhu}},
  \bibinfo {author} {\bibfnamefont {A.}~\bibnamefont {Sen(De)}}, \ and\
  \bibinfo {author} {\bibfnamefont {U.}~\bibnamefont {Sen}},\ }\href {\doibase
  10.1103/PhysRevA.90.022319} {\bibfield  {journal} {\bibinfo  {journal} {Phys.
  Rev. A}\ }\textbf {\bibinfo {volume} {90}},\ \bibinfo {pages} {022319}
  (\bibinfo {year} {2014})}\BibitemShut {NoStop}%
\bibitem [{\citenamefont {Ekert}(1991)}]{Ekert_1991}%
  \BibitemOpen
  \bibfield  {author} {\bibinfo {author} {\bibfnamefont {A.~K.}\ \bibnamefont
  {Ekert}},\ }\href {\doibase 10.1103/PhysRevLett.67.661} {\bibfield  {journal}
  {\bibinfo  {journal} {Phys. Rev. Lett.}\ }\textbf {\bibinfo {volume} {67}},\
  \bibinfo {pages} {661} (\bibinfo {year} {1991})}\BibitemShut {NoStop}%
\bibitem [{\citenamefont {Hillery}\ \emph {et~al.}(1999)\citenamefont
  {Hillery}, \citenamefont {Bu\ifmmode~\check{z}\else \v{z}\fi{}ek},\ and\
  \citenamefont {Berthiaume}}]{Hillery_1999}%
  \BibitemOpen
  \bibfield  {author} {\bibinfo {author} {\bibfnamefont {M.}~\bibnamefont
  {Hillery}}, \bibinfo {author} {\bibfnamefont {V.}~\bibnamefont
  {Bu\ifmmode~\check{z}\else \v{z}\fi{}ek}}, \ and\ \bibinfo {author}
  {\bibfnamefont {A.}~\bibnamefont {Berthiaume}},\ }\href {\doibase
  10.1103/PhysRevA.59.1829} {\bibfield  {journal} {\bibinfo  {journal} {Phys.
  Rev. A}\ }\textbf {\bibinfo {volume} {59}},\ \bibinfo {pages} {1829}
  (\bibinfo {year} {1999})}\BibitemShut {NoStop}%
\bibitem [{\citenamefont {Shor}\ and\ \citenamefont
  {Preskill}(2000)}]{Shor_2000}%
  \BibitemOpen
  \bibfield  {author} {\bibinfo {author} {\bibfnamefont {P.~W.}\ \bibnamefont
  {Shor}}\ and\ \bibinfo {author} {\bibfnamefont {J.}~\bibnamefont
  {Preskill}},\ }\href {\doibase 10.1103/PhysRevLett.85.441} {\bibfield
  {journal} {\bibinfo  {journal} {Phys. Rev. Lett.}\ }\textbf {\bibinfo
  {volume} {85}},\ \bibinfo {pages} {441} (\bibinfo {year} {2000})}\BibitemShut
  {NoStop}%
\bibitem [{\citenamefont {Adhikari}\ \emph {et~al.}(2010)\citenamefont
  {Adhikari}, \citenamefont {Chakrabarty},\ and\ \citenamefont
  {Agrawal}}]{Adhikari_2010}%
  \BibitemOpen
  \bibfield  {author} {\bibinfo {author} {\bibfnamefont {S.}~\bibnamefont
  {Adhikari}}, \bibinfo {author} {\bibfnamefont {I.}~\bibnamefont
  {Chakrabarty}}, \ and\ \bibinfo {author} {\bibfnamefont {P.}~\bibnamefont
  {Agrawal}},\ }\href {\doibase 10.26421/QIC12.3-4-5} {\bibfield  {journal}
  {\bibinfo  {journal} {Quantum Information and Computation}\ }\textbf
  {\bibinfo {volume} {12}} (\bibinfo {year} {2010}),\
  10.26421/QIC12.3-4-5}\BibitemShut {NoStop}%
\bibitem [{\citenamefont {Bennett}\ and\ \citenamefont
  {Brassard}(2014)}]{Bennett_2014}%
  \BibitemOpen
  \bibfield  {author} {\bibinfo {author} {\bibfnamefont {C.~H.}\ \bibnamefont
  {Bennett}}\ and\ \bibinfo {author} {\bibfnamefont {G.}~\bibnamefont
  {Brassard}},\ }\href {\doibase 10.1016/j.tcs.2014.05.025} {\bibfield
  {journal} {\bibinfo  {journal} {Theoretical Computer Science}\ }\textbf
  {\bibinfo {volume} {560}},\ \bibinfo {pages} {7} (\bibinfo {year}
  {2014})}\BibitemShut {NoStop}%
\bibitem [{\citenamefont {Sazim}\ \emph {et~al.}(2015)\citenamefont {Sazim},
  \citenamefont {Chiranjeevi}, \citenamefont {Chakrabarty},\ and\ \citenamefont
  {Srinathan}}]{Sazim_2015}%
  \BibitemOpen
  \bibfield  {author} {\bibinfo {author} {\bibfnamefont {S.}~\bibnamefont
  {Sazim}}, \bibinfo {author} {\bibfnamefont {V.}~\bibnamefont {Chiranjeevi}},
  \bibinfo {author} {\bibfnamefont {I.}~\bibnamefont {Chakrabarty}}, \ and\
  \bibinfo {author} {\bibfnamefont {K.}~\bibnamefont {Srinathan}},\ }\href
  {\doibase 10.1007/s11128-015-1109-7} {\bibfield  {journal} {\bibinfo
  {journal} {Quantum Information Processing}\ }\textbf {\bibinfo {volume}
  {14}},\ \bibinfo {pages} {4651} (\bibinfo {year} {2015})}\BibitemShut
  {NoStop}%
\bibitem [{\citenamefont {Ray}\ \emph {et~al.}(2016)\citenamefont {Ray},
  \citenamefont {Chatterjee},\ and\ \citenamefont {Chakrabarty}}]{Ray_2016}%
  \BibitemOpen
  \bibfield  {author} {\bibinfo {author} {\bibfnamefont {M.}~\bibnamefont
  {Ray}}, \bibinfo {author} {\bibfnamefont {S.}~\bibnamefont {Chatterjee}}, \
  and\ \bibinfo {author} {\bibfnamefont {I.}~\bibnamefont {Chakrabarty}},\
  }\href {\doibase 10.1140/epjd/e2016-60683-x} {\bibfield  {journal} {\bibinfo
  {journal} {The European Physical Journal D}\ }\textbf {\bibinfo {volume}
  {70}},\ \bibinfo {pages} {114} (\bibinfo {year} {2016})}\BibitemShut
  {NoStop}%
\bibitem [{\citenamefont {Mudholkar}\ \emph {et~al.}(2023)\citenamefont
  {Mudholkar}, \citenamefont {Vanarasa}, \citenamefont {Chakrabarty},\ and\
  \citenamefont {Kannan}}]{mudholkar_2023}%
  \BibitemOpen
  \bibfield  {author} {\bibinfo {author} {\bibfnamefont {P.}~\bibnamefont
  {Mudholkar}}, \bibinfo {author} {\bibfnamefont {C.}~\bibnamefont {Vanarasa}},
  \bibinfo {author} {\bibfnamefont {I.}~\bibnamefont {Chakrabarty}}, \ and\
  \bibinfo {author} {\bibfnamefont {S.}~\bibnamefont {Kannan}},\ }\href@noop {}
  {\enquote {\bibinfo {title} {Revocation and reconstruction of shared quantum
  secrets},}\ } (\bibinfo {year} {2023}),\ \Eprint
  {http://arxiv.org/abs/2112.15556} {arXiv:2112.15556 [quant-ph]} \BibitemShut
  {NoStop}%
\bibitem [{\citenamefont {Kimble}(2008)}]{Kimble2008}%
  \BibitemOpen
  \bibfield  {author} {\bibinfo {author} {\bibfnamefont {H.~J.}\ \bibnamefont
  {Kimble}},\ }\href {\doibase 10.1038/nature07127} {\bibfield  {journal}
  {\bibinfo  {journal} {Nature}\ }\textbf {\bibinfo {volume} {453}},\ \bibinfo
  {pages} {1023} (\bibinfo {year} {2008})}\BibitemShut {NoStop}%
\bibitem [{\citenamefont {Wehner}\ \emph {et~al.}(2018)\citenamefont {Wehner},
  \citenamefont {Elkouss},\ and\ \citenamefont {Hanson}}]{wehner_18internet}%
  \BibitemOpen
  \bibfield  {author} {\bibinfo {author} {\bibfnamefont {S.}~\bibnamefont
  {Wehner}}, \bibinfo {author} {\bibfnamefont {D.}~\bibnamefont {Elkouss}}, \
  and\ \bibinfo {author} {\bibfnamefont {R.}~\bibnamefont {Hanson}},\ }\href
  {\doibase 10.1126/science.aam9288} {\bibfield  {journal} {\bibinfo  {journal}
  {Science}\ }\textbf {\bibinfo {volume} {362}},\ \bibinfo {pages} {eaam9288}
  (\bibinfo {year} {2018})},\ \Eprint
  {http://arxiv.org/abs/https://www.science.org/doi/pdf/10.1126/science.aam9288}
  {https://www.science.org/doi/pdf/10.1126/science.aam9288} \BibitemShut
  {NoStop}%
\bibitem [{\citenamefont {Pirker}\ and\ \citenamefont
  {Dür}(2019)}]{Pirker2019}%
  \BibitemOpen
  \bibfield  {author} {\bibinfo {author} {\bibfnamefont {A.}~\bibnamefont
  {Pirker}}\ and\ \bibinfo {author} {\bibfnamefont {W.}~\bibnamefont {Dür}},\
  }\href {\doibase 10.1088/1367-2630/ab05f7} {\bibfield  {journal} {\bibinfo
  {journal} {New Journal of Physics}\ }\textbf {\bibinfo {volume} {21}},\
  \bibinfo {pages} {033003} (\bibinfo {year} {2019})}\BibitemShut {NoStop}%
\bibitem [{\citenamefont {Wei}\ \emph {et~al.}(2022)\citenamefont {Wei},
  \citenamefont {Jing}, \citenamefont {Zhang}, \citenamefont {Liao},
  \citenamefont {Yuan}, \citenamefont {Fan}, \citenamefont {Lyu}, \citenamefont
  {Zhou}, \citenamefont {Wang}, \citenamefont {Deng}, \citenamefont {Song},
  \citenamefont {Oblak}, \citenamefont {Guo},\ and\ \citenamefont
  {Zhou}}]{Wei2022}%
  \BibitemOpen
  \bibfield  {author} {\bibinfo {author} {\bibfnamefont {S.-H.}\ \bibnamefont
  {Wei}}, \bibinfo {author} {\bibfnamefont {B.}~\bibnamefont {Jing}}, \bibinfo
  {author} {\bibfnamefont {X.-Y.}\ \bibnamefont {Zhang}}, \bibinfo {author}
  {\bibfnamefont {J.-Y.}\ \bibnamefont {Liao}}, \bibinfo {author}
  {\bibfnamefont {C.-Z.}\ \bibnamefont {Yuan}}, \bibinfo {author}
  {\bibfnamefont {B.-Y.}\ \bibnamefont {Fan}}, \bibinfo {author} {\bibfnamefont
  {C.}~\bibnamefont {Lyu}}, \bibinfo {author} {\bibfnamefont {D.-L.}\
  \bibnamefont {Zhou}}, \bibinfo {author} {\bibfnamefont {Y.}~\bibnamefont
  {Wang}}, \bibinfo {author} {\bibfnamefont {G.-W.}\ \bibnamefont {Deng}},
  \bibinfo {author} {\bibfnamefont {H.-Z.}\ \bibnamefont {Song}}, \bibinfo
  {author} {\bibfnamefont {D.}~\bibnamefont {Oblak}}, \bibinfo {author}
  {\bibfnamefont {G.-C.}\ \bibnamefont {Guo}}, \ and\ \bibinfo {author}
  {\bibfnamefont {Q.}~\bibnamefont {Zhou}},\ }\href {\doibase
  https://doi.org/10.1002/lpor.202100219} {\bibfield  {journal} {\bibinfo
  {journal} {Laser \& Photonics Reviews}\ }\textbf {\bibinfo {volume} {16}},\
  \bibinfo {pages} {2100219} (\bibinfo {year} {2022})}\BibitemShut {NoStop}%
\bibitem [{\citenamefont {Azuma}\ \emph {et~al.}(2023)\citenamefont {Azuma},
  \citenamefont {Economou}, \citenamefont {Elkouss}, \citenamefont {Hilaire},
  \citenamefont {Jiang}, \citenamefont {Lo},\ and\ \citenamefont
  {Tzitrin}}]{azuma2023quantum}%
  \BibitemOpen
  \bibfield  {author} {\bibinfo {author} {\bibfnamefont {K.}~\bibnamefont
  {Azuma}}, \bibinfo {author} {\bibfnamefont {S.~E.}\ \bibnamefont {Economou}},
  \bibinfo {author} {\bibfnamefont {D.}~\bibnamefont {Elkouss}}, \bibinfo
  {author} {\bibfnamefont {P.}~\bibnamefont {Hilaire}}, \bibinfo {author}
  {\bibfnamefont {L.}~\bibnamefont {Jiang}}, \bibinfo {author} {\bibfnamefont
  {H.-K.}\ \bibnamefont {Lo}}, \ and\ \bibinfo {author} {\bibfnamefont
  {I.}~\bibnamefont {Tzitrin}},\ }\href@noop {} {\enquote {\bibinfo {title}
  {Quantum repeaters: From quantum networks to the quantum internet},}\ }
  (\bibinfo {year} {2023}),\ \Eprint {http://arxiv.org/abs/2212.10820}
  {arXiv:2212.10820 [quant-ph]} \BibitemShut {NoStop}%
\bibitem [{\citenamefont {Raussendorf}\ and\ \citenamefont
  {Briegel}(2001)}]{Raussendorf_2001}%
  \BibitemOpen
  \bibfield  {author} {\bibinfo {author} {\bibfnamefont {R.}~\bibnamefont
  {Raussendorf}}\ and\ \bibinfo {author} {\bibfnamefont {H.~J.}\ \bibnamefont
  {Briegel}},\ }\href {\doibase 10.1103/PhysRevLett.86.5188} {\bibfield
  {journal} {\bibinfo  {journal} {Phys. Rev. Lett.}\ }\textbf {\bibinfo
  {volume} {86}},\ \bibinfo {pages} {5188} (\bibinfo {year}
  {2001})}\BibitemShut {NoStop}%
\bibitem [{\citenamefont {Hein}\ \emph {et~al.}(2004)\citenamefont {Hein},
  \citenamefont {Eisert},\ and\ \citenamefont {Briegel}}]{Hein_2004}%
  \BibitemOpen
  \bibfield  {author} {\bibinfo {author} {\bibfnamefont {M.}~\bibnamefont
  {Hein}}, \bibinfo {author} {\bibfnamefont {J.}~\bibnamefont {Eisert}}, \ and\
  \bibinfo {author} {\bibfnamefont {H.~J.}\ \bibnamefont {Briegel}},\ }\href
  {\doibase 10.1103/PhysRevA.69.062311} {\bibfield  {journal} {\bibinfo
  {journal} {Phys. Rev. A}\ }\textbf {\bibinfo {volume} {69}},\ \bibinfo
  {pages} {062311} (\bibinfo {year} {2004})}\BibitemShut {NoStop}%
\bibitem [{\citenamefont {Hein}\ \emph {et~al.}(2006)\citenamefont {Hein},
  \citenamefont {Dür}, \citenamefont {Eisert}, \citenamefont {Raussendorf},
  \citenamefont {den Nest},\ and\ \citenamefont {Briegel}}]{Hein_2006}%
  \BibitemOpen
  \bibfield  {author} {\bibinfo {author} {\bibfnamefont {M.}~\bibnamefont
  {Hein}}, \bibinfo {author} {\bibfnamefont {W.}~\bibnamefont {Dür}}, \bibinfo
  {author} {\bibfnamefont {J.}~\bibnamefont {Eisert}}, \bibinfo {author}
  {\bibfnamefont {R.}~\bibnamefont {Raussendorf}}, \bibinfo {author}
  {\bibfnamefont {M.~V.}\ \bibnamefont {den Nest}}, \ and\ \bibinfo {author}
  {\bibfnamefont {H.~J.}\ \bibnamefont {Briegel}},\ }\href@noop {} {\enquote
  {\bibinfo {title} {Entanglement in graph states and its applications},}\ }
  (\bibinfo {year} {2006}),\ \Eprint {http://arxiv.org/abs/quant-ph/0602096}
  {arXiv:quant-ph/0602096 [quant-ph]} \BibitemShut {NoStop}%
\bibitem [{\citenamefont {Briegel}\ \emph
  {et~al.}(2009{\natexlab{a}})\citenamefont {Briegel}, \citenamefont {Browne},
  \citenamefont {Dür}, \citenamefont {Raussendorf},\ and\ \citenamefont {den
  Nest}}]{Briegel_2009}%
  \BibitemOpen
  \bibfield  {author} {\bibinfo {author} {\bibfnamefont {H.~J.}\ \bibnamefont
  {Briegel}}, \bibinfo {author} {\bibfnamefont {D.~E.}\ \bibnamefont {Browne}},
  \bibinfo {author} {\bibfnamefont {W.}~\bibnamefont {Dür}}, \bibinfo {author}
  {\bibfnamefont {R.}~\bibnamefont {Raussendorf}}, \ and\ \bibinfo {author}
  {\bibfnamefont {M.~V.}\ \bibnamefont {den Nest}},\ }\href {\doibase
  10.1038/nphys1157} {\bibfield  {journal} {\bibinfo  {journal} {Nature
  Physics}\ }\textbf {\bibinfo {volume} {5}},\ \bibinfo {pages} {19} (\bibinfo
  {year} {2009}{\natexlab{a}})}\BibitemShut {NoStop}%
\bibitem [{\citenamefont {Beals}\ \emph {et~al.}(2013)\citenamefont {Beals},
  \citenamefont {Brierley}, \citenamefont {Gray}, \citenamefont {Harrow},
  \citenamefont {Kutin}, \citenamefont {Linden}, \citenamefont {Shepherd},\
  and\ \citenamefont {Stather}}]{Beals_2013}%
  \BibitemOpen
  \bibfield  {author} {\bibinfo {author} {\bibfnamefont {R.}~\bibnamefont
  {Beals}}, \bibinfo {author} {\bibfnamefont {S.}~\bibnamefont {Brierley}},
  \bibinfo {author} {\bibfnamefont {O.}~\bibnamefont {Gray}}, \bibinfo {author}
  {\bibfnamefont {A.~W.}\ \bibnamefont {Harrow}}, \bibinfo {author}
  {\bibfnamefont {S.}~\bibnamefont {Kutin}}, \bibinfo {author} {\bibfnamefont
  {N.}~\bibnamefont {Linden}}, \bibinfo {author} {\bibfnamefont
  {D.}~\bibnamefont {Shepherd}}, \ and\ \bibinfo {author} {\bibfnamefont
  {M.}~\bibnamefont {Stather}},\ }\href {\doibase 10.1098/rspa.2012.0686}
  {\bibfield  {journal} {\bibinfo  {journal} {Proceedings of the Royal Society
  A: Mathematical, Physical and Engineering Sciences}\ }\textbf {\bibinfo
  {volume} {469}},\ \bibinfo {pages} {20120686} (\bibinfo {year}
  {2013})}\BibitemShut {NoStop}%
\bibitem [{\citenamefont {Cleve}\ \emph {et~al.}(1999)\citenamefont {Cleve},
  \citenamefont {Gottesman},\ and\ \citenamefont {Lo}}]{Cleve_1999}%
  \BibitemOpen
  \bibfield  {author} {\bibinfo {author} {\bibfnamefont {R.}~\bibnamefont
  {Cleve}}, \bibinfo {author} {\bibfnamefont {D.}~\bibnamefont {Gottesman}}, \
  and\ \bibinfo {author} {\bibfnamefont {H.-K.}\ \bibnamefont {Lo}},\ }\href
  {\doibase 10.1103/PhysRevLett.83.648} {\bibfield  {journal} {\bibinfo
  {journal} {Phys. Rev. Lett.}\ }\textbf {\bibinfo {volume} {83}},\ \bibinfo
  {pages} {648} (\bibinfo {year} {1999})}\BibitemShut {NoStop}%
\bibitem [{\citenamefont {Karlsson}\ \emph {et~al.}(1999)\citenamefont
  {Karlsson}, \citenamefont {Koashi},\ and\ \citenamefont
  {Imoto}}]{Karlsson_1999}%
  \BibitemOpen
  \bibfield  {author} {\bibinfo {author} {\bibfnamefont {A.}~\bibnamefont
  {Karlsson}}, \bibinfo {author} {\bibfnamefont {M.}~\bibnamefont {Koashi}}, \
  and\ \bibinfo {author} {\bibfnamefont {N.}~\bibnamefont {Imoto}},\ }\href
  {\doibase 10.1103/PhysRevA.59.162} {\bibfield  {journal} {\bibinfo  {journal}
  {Phys. Rev. A}\ }\textbf {\bibinfo {volume} {59}},\ \bibinfo {pages} {162}
  (\bibinfo {year} {1999})}\BibitemShut {NoStop}%
\bibitem [{\citenamefont {Jennewein}\ \emph {et~al.}(2000)\citenamefont
  {Jennewein}, \citenamefont {Simon}, \citenamefont {Weihs}, \citenamefont
  {Weinfurter},\ and\ \citenamefont {Zeilinger}}]{Jennewein_2000}%
  \BibitemOpen
  \bibfield  {author} {\bibinfo {author} {\bibfnamefont {T.}~\bibnamefont
  {Jennewein}}, \bibinfo {author} {\bibfnamefont {C.}~\bibnamefont {Simon}},
  \bibinfo {author} {\bibfnamefont {G.}~\bibnamefont {Weihs}}, \bibinfo
  {author} {\bibfnamefont {H.}~\bibnamefont {Weinfurter}}, \ and\ \bibinfo
  {author} {\bibfnamefont {A.}~\bibnamefont {Zeilinger}},\ }\href {\doibase
  10.1103/PhysRevLett.84.4729} {\bibfield  {journal} {\bibinfo  {journal}
  {Phys. Rev. Lett.}\ }\textbf {\bibinfo {volume} {84}},\ \bibinfo {pages}
  {4729} (\bibinfo {year} {2000})}\BibitemShut {NoStop}%
\bibitem [{\citenamefont {Gisin}\ \emph {et~al.}(2002)\citenamefont {Gisin},
  \citenamefont {Ribordy}, \citenamefont {Tittel},\ and\ \citenamefont
  {Zbinden}}]{Gisin_2002}%
  \BibitemOpen
  \bibfield  {author} {\bibinfo {author} {\bibfnamefont {N.}~\bibnamefont
  {Gisin}}, \bibinfo {author} {\bibfnamefont {G.}~\bibnamefont {Ribordy}},
  \bibinfo {author} {\bibfnamefont {W.}~\bibnamefont {Tittel}}, \ and\ \bibinfo
  {author} {\bibfnamefont {H.}~\bibnamefont {Zbinden}},\ }\href {\doibase
  10.1103/RevModPhys.74.145} {\bibfield  {journal} {\bibinfo  {journal} {Rev.
  Mod. Phys.}\ }\textbf {\bibinfo {volume} {74}},\ \bibinfo {pages} {145}
  (\bibinfo {year} {2002})}\BibitemShut {NoStop}%
\bibitem [{\citenamefont {Hillery}\ \emph {et~al.}(2006)\citenamefont
  {Hillery}, \citenamefont {Ziman}, \citenamefont {Bu{\v{z}}ek},\ and\
  \citenamefont {Bielikov{\'{a}}}}]{Hillery_2006}%
  \BibitemOpen
  \bibfield  {author} {\bibinfo {author} {\bibfnamefont {M.}~\bibnamefont
  {Hillery}}, \bibinfo {author} {\bibfnamefont {M.}~\bibnamefont {Ziman}},
  \bibinfo {author} {\bibfnamefont {V.}~\bibnamefont {Bu{\v{z}}ek}}, \ and\
  \bibinfo {author} {\bibfnamefont {M.}~\bibnamefont {Bielikov{\'{a}}}},\
  }\href {\doibase 10.1016/j.physleta.2005.09.010} {\bibfield  {journal}
  {\bibinfo  {journal} {Physics Letters A}\ }\textbf {\bibinfo {volume}
  {349}},\ \bibinfo {pages} {75} (\bibinfo {year} {2006})}\BibitemShut
  {NoStop}%
\bibitem [{\citenamefont {Xu}\ \emph {et~al.}(2014)\citenamefont {Xu},
  \citenamefont {Wen}, \citenamefont {Gao},\ and\ \citenamefont
  {Qin}}]{Xu_2014}%
  \BibitemOpen
  \bibfield  {author} {\bibinfo {author} {\bibfnamefont {G.-B.}\ \bibnamefont
  {Xu}}, \bibinfo {author} {\bibfnamefont {Q.-Y.}\ \bibnamefont {Wen}},
  \bibinfo {author} {\bibfnamefont {F.}~\bibnamefont {Gao}}, \ and\ \bibinfo
  {author} {\bibfnamefont {S.-J.}\ \bibnamefont {Qin}},\ }\href {\doibase
  10.1007/s11128-014-0816-9} {\bibfield  {journal} {\bibinfo  {journal}
  {Quantum Information Processing}\ }\textbf {\bibinfo {volume} {13}},\
  \bibinfo {pages} {2587} (\bibinfo {year} {2014})}\BibitemShut {NoStop}%
\bibitem [{\citenamefont {Nielsen}\ and\ \citenamefont
  {Chuang}(2010)}]{Nielsen_2010}%
  \BibitemOpen
  \bibfield  {author} {\bibinfo {author} {\bibfnamefont {M.~A.}\ \bibnamefont
  {Nielsen}}\ and\ \bibinfo {author} {\bibfnamefont {I.~L.}\ \bibnamefont
  {Chuang}},\ }\href {\doibase DOI: 10.1017/CBO9780511976667} {\emph {\bibinfo
  {title} {Quantum Computation and Quantum Information: 10th Anniversary
  Edition}}}\ (\bibinfo  {publisher} {Cambridge University Press},\ \bibinfo
  {year} {2010})\BibitemShut {NoStop}%
\bibitem [{\citenamefont {Walther}\ \emph {et~al.}(2005)\citenamefont
  {Walther}, \citenamefont {Resch}, \citenamefont {Rudolph}, \citenamefont
  {Schenck}, \citenamefont {Weinfurter}, \citenamefont {Vedral}, \citenamefont
  {Aspelmeyer},\ and\ \citenamefont {Zeilinger}}]{Walther_2005}%
  \BibitemOpen
  \bibfield  {author} {\bibinfo {author} {\bibfnamefont {P.}~\bibnamefont
  {Walther}}, \bibinfo {author} {\bibfnamefont {K.}~\bibnamefont {Resch}},
  \bibinfo {author} {\bibfnamefont {T.}~\bibnamefont {Rudolph}}, \bibinfo
  {author} {\bibfnamefont {E.}~\bibnamefont {Schenck}}, \bibinfo {author}
  {\bibfnamefont {H.}~\bibnamefont {Weinfurter}}, \bibinfo {author}
  {\bibfnamefont {V.}~\bibnamefont {Vedral}}, \bibinfo {author} {\bibfnamefont
  {M.}~\bibnamefont {Aspelmeyer}}, \ and\ \bibinfo {author} {\bibfnamefont
  {A.}~\bibnamefont {Zeilinger}},\ }\href {\doibase 10.1038/nature03347}
  {\bibfield  {journal} {\bibinfo  {journal} {Nature}\ }\textbf {\bibinfo
  {volume} {434}},\ \bibinfo {pages} {169} (\bibinfo {year}
  {2005})}\BibitemShut {NoStop}%
\bibitem [{\citenamefont {Acin}\ \emph {et~al.}(2007)\citenamefont {Acin},
  \citenamefont {Cirac},\ and\ \citenamefont {Lewenstein}}]{Acin_2007}%
  \BibitemOpen
  \bibfield  {author} {\bibinfo {author} {\bibfnamefont {A.}~\bibnamefont
  {Acin}}, \bibinfo {author} {\bibfnamefont {J.~I.}\ \bibnamefont {Cirac}}, \
  and\ \bibinfo {author} {\bibfnamefont {M.}~\bibnamefont {Lewenstein}},\
  }\href {\doibase 10.1038/nphys549} {\bibfield  {journal} {\bibinfo  {journal}
  {Nature Physics}\ }\textbf {\bibinfo {volume} {3}},\ \bibinfo {pages} {256}
  (\bibinfo {year} {2007})}\BibitemShut {NoStop}%
\bibitem [{\citenamefont {Briegel}\ \emph
  {et~al.}(2009{\natexlab{b}})\citenamefont {Briegel}, \citenamefont {Browne},
  \citenamefont {Dür}, \citenamefont {Raussendorf},\ and\ \citenamefont {den
  Nest}}]{Briegel2009}%
  \BibitemOpen
  \bibfield  {author} {\bibinfo {author} {\bibfnamefont {H.~J.}\ \bibnamefont
  {Briegel}}, \bibinfo {author} {\bibfnamefont {D.~E.}\ \bibnamefont {Browne}},
  \bibinfo {author} {\bibfnamefont {W.}~\bibnamefont {Dür}}, \bibinfo {author}
  {\bibfnamefont {R.}~\bibnamefont {Raussendorf}}, \ and\ \bibinfo {author}
  {\bibfnamefont {M.~V.}\ \bibnamefont {den Nest}},\ }\href {\doibase
  10.1038/nphys1157} {\bibfield  {journal} {\bibinfo  {journal} {Nature
  Physics}\ }\textbf {\bibinfo {volume} {5}},\ \bibinfo {pages} {19} (\bibinfo
  {year} {2009}{\natexlab{b}})}\BibitemShut {NoStop}%
\bibitem [{\citenamefont {Reiserer}(2022)}]{reiserer22}%
  \BibitemOpen
  \bibfield  {author} {\bibinfo {author} {\bibfnamefont {A.}~\bibnamefont
  {Reiserer}},\ }\href {\doibase 10.1103/RevModPhys.94.041003} {\bibfield
  {journal} {\bibinfo  {journal} {Rev. Mod. Phys.}\ }\textbf {\bibinfo {volume}
  {94}},\ \bibinfo {pages} {041003} (\bibinfo {year} {2022})}\BibitemShut
  {NoStop}%
\bibitem [{\citenamefont {Pan}\ \emph {et~al.}(2012)\citenamefont {Pan},
  \citenamefont {Chen}, \citenamefont {Lu}, \citenamefont {Weinfurter},
  \citenamefont {Zeilinger},\ and\ \citenamefont {\ifmmode~\dot{Z}\else
  \.{Z}\fi{}ukowski}}]{pan12}%
  \BibitemOpen
  \bibfield  {author} {\bibinfo {author} {\bibfnamefont {J.-W.}\ \bibnamefont
  {Pan}}, \bibinfo {author} {\bibfnamefont {Z.-B.}\ \bibnamefont {Chen}},
  \bibinfo {author} {\bibfnamefont {C.-Y.}\ \bibnamefont {Lu}}, \bibinfo
  {author} {\bibfnamefont {H.}~\bibnamefont {Weinfurter}}, \bibinfo {author}
  {\bibfnamefont {A.}~\bibnamefont {Zeilinger}}, \ and\ \bibinfo {author}
  {\bibfnamefont {M.}~\bibnamefont {\ifmmode~\dot{Z}\else \.{Z}\fi{}ukowski}},\
  }\href {\doibase 10.1103/RevModPhys.84.777} {\bibfield  {journal} {\bibinfo
  {journal} {Rev. Mod. Phys.}\ }\textbf {\bibinfo {volume} {84}},\ \bibinfo
  {pages} {777} (\bibinfo {year} {2012})}\BibitemShut {NoStop}%
\bibitem [{\citenamefont {Duan}\ and\ \citenamefont {Monroe}(2010)}]{duan10}%
  \BibitemOpen
  \bibfield  {author} {\bibinfo {author} {\bibfnamefont {L.-M.}\ \bibnamefont
  {Duan}}\ and\ \bibinfo {author} {\bibfnamefont {C.}~\bibnamefont {Monroe}},\
  }\href {\doibase 10.1103/RevModPhys.82.1209} {\bibfield  {journal} {\bibinfo
  {journal} {Rev. Mod. Phys.}\ }\textbf {\bibinfo {volume} {82}},\ \bibinfo
  {pages} {1209} (\bibinfo {year} {2010})}\BibitemShut {NoStop}%
\bibitem [{\citenamefont {Monroe}\ \emph {et~al.}(2013)\citenamefont {Monroe},
  \citenamefont {Campbell}, \citenamefont {Cao}, \citenamefont {Choi},
  \citenamefont {Clark}, \citenamefont {Debnath}, \citenamefont {Figgatt},
  \citenamefont {Hayes}, \citenamefont {Hucul}, \citenamefont {Inlek},
  \citenamefont {Islam}, \citenamefont {Korenblit}, \citenamefont {Johnson},
  \citenamefont {Manning}, \citenamefont {Mizrahi}, \citenamefont {Neyenhuis},
  \citenamefont {Lee}, \citenamefont {Richerme}, \citenamefont {Senko},
  \citenamefont {Smith},\ and\ \citenamefont {Wright}}]{Monroe2013}%
  \BibitemOpen
  \bibfield  {author} {\bibinfo {author} {\bibfnamefont {C.}~\bibnamefont
  {Monroe}}, \bibinfo {author} {\bibfnamefont {W.}~\bibnamefont {Campbell}},
  \bibinfo {author} {\bibfnamefont {C.}~\bibnamefont {Cao}}, \bibinfo {author}
  {\bibfnamefont {T.}~\bibnamefont {Choi}}, \bibinfo {author} {\bibfnamefont
  {S.}~\bibnamefont {Clark}}, \bibinfo {author} {\bibfnamefont
  {S.}~\bibnamefont {Debnath}}, \bibinfo {author} {\bibfnamefont
  {C.}~\bibnamefont {Figgatt}}, \bibinfo {author} {\bibfnamefont
  {D.}~\bibnamefont {Hayes}}, \bibinfo {author} {\bibfnamefont
  {D.}~\bibnamefont {Hucul}}, \bibinfo {author} {\bibfnamefont
  {V.}~\bibnamefont {Inlek}}, \bibinfo {author} {\bibfnamefont
  {R.}~\bibnamefont {Islam}}, \bibinfo {author} {\bibfnamefont
  {S.}~\bibnamefont {Korenblit}}, \bibinfo {author} {\bibfnamefont
  {K.}~\bibnamefont {Johnson}}, \bibinfo {author} {\bibfnamefont
  {A.}~\bibnamefont {Manning}}, \bibinfo {author} {\bibfnamefont
  {J.}~\bibnamefont {Mizrahi}}, \bibinfo {author} {\bibfnamefont
  {B.}~\bibnamefont {Neyenhuis}}, \bibinfo {author} {\bibfnamefont
  {A.}~\bibnamefont {Lee}}, \bibinfo {author} {\bibfnamefont {P.}~\bibnamefont
  {Richerme}}, \bibinfo {author} {\bibfnamefont {C.}~\bibnamefont {Senko}},
  \bibinfo {author} {\bibfnamefont {J.}~\bibnamefont {Smith}}, \ and\ \bibinfo
  {author} {\bibfnamefont {K.}~\bibnamefont {Wright}},\ }\href {\doibase
  10.1088/1742-6596/467/1/012008} {\bibfield  {journal} {\bibinfo  {journal}
  {Journal of Physics: Conference Series}\ }\textbf {\bibinfo {volume} {467}},\
  \bibinfo {pages} {012008} (\bibinfo {year} {2013})}\BibitemShut {NoStop}%
\bibitem [{\citenamefont {Hermans}\ \emph {et~al.}(2022)\citenamefont
  {Hermans}, \citenamefont {Pompili}, \citenamefont {Beukers}, \citenamefont
  {Baier}, \citenamefont {Borregaard},\ and\ \citenamefont
  {Hanson}}]{Hermans2022}%
  \BibitemOpen
  \bibfield  {author} {\bibinfo {author} {\bibfnamefont {S.~L.~N.}\
  \bibnamefont {Hermans}}, \bibinfo {author} {\bibfnamefont {M.}~\bibnamefont
  {Pompili}}, \bibinfo {author} {\bibfnamefont {H.~K.~C.}\ \bibnamefont
  {Beukers}}, \bibinfo {author} {\bibfnamefont {S.}~\bibnamefont {Baier}},
  \bibinfo {author} {\bibfnamefont {J.}~\bibnamefont {Borregaard}}, \ and\
  \bibinfo {author} {\bibfnamefont {R.}~\bibnamefont {Hanson}},\ }\href
  {\doibase 10.1038/s41586-022-04697-y} {\bibfield  {journal} {\bibinfo
  {journal} {Nature}\ }\textbf {\bibinfo {volume} {605}},\ \bibinfo {pages}
  {663} (\bibinfo {year} {2022})}\BibitemShut {NoStop}%
\bibitem [{\citenamefont {Gao}\ \emph {et~al.}(2015)\citenamefont {Gao},
  \citenamefont {Imamoglu}, \citenamefont {Bernien},\ and\ \citenamefont
  {Hanson}}]{Gao2015}%
  \BibitemOpen
  \bibfield  {author} {\bibinfo {author} {\bibfnamefont {W.~B.}\ \bibnamefont
  {Gao}}, \bibinfo {author} {\bibfnamefont {A.}~\bibnamefont {Imamoglu}},
  \bibinfo {author} {\bibfnamefont {H.}~\bibnamefont {Bernien}}, \ and\
  \bibinfo {author} {\bibfnamefont {R.}~\bibnamefont {Hanson}},\ }\href
  {\doibase 10.1038/nphoton.2015.58} {\bibfield  {journal} {\bibinfo  {journal}
  {Nature Photonics}\ }\textbf {\bibinfo {volume} {9}},\ \bibinfo {pages} {363}
  (\bibinfo {year} {2015})}\BibitemShut {NoStop}%
\bibitem [{\citenamefont {Briegel}\ \emph {et~al.}(1998)\citenamefont
  {Briegel}, \citenamefont {D\"ur}, \citenamefont {Cirac},\ and\ \citenamefont
  {Zoller}}]{briegel_repeater_98}%
  \BibitemOpen
  \bibfield  {author} {\bibinfo {author} {\bibfnamefont {H.-J.}\ \bibnamefont
  {Briegel}}, \bibinfo {author} {\bibfnamefont {W.}~\bibnamefont {D\"ur}},
  \bibinfo {author} {\bibfnamefont {J.~I.}\ \bibnamefont {Cirac}}, \ and\
  \bibinfo {author} {\bibfnamefont {P.}~\bibnamefont {Zoller}},\ }\href
  {\doibase 10.1103/PhysRevLett.81.5932} {\bibfield  {journal} {\bibinfo
  {journal} {Phys. Rev. Lett.}\ }\textbf {\bibinfo {volume} {81}},\ \bibinfo
  {pages} {5932} (\bibinfo {year} {1998})}\BibitemShut {NoStop}%
\bibitem [{\citenamefont {Ladd}\ \emph {et~al.}(2006)\citenamefont {Ladd},
  \citenamefont {van Loock}, \citenamefont {Nemoto}, \citenamefont {Munro},\
  and\ \citenamefont {Yamamoto}}]{Ladd_2006}%
  \BibitemOpen
  \bibfield  {author} {\bibinfo {author} {\bibfnamefont {T.~D.}\ \bibnamefont
  {Ladd}}, \bibinfo {author} {\bibfnamefont {P.}~\bibnamefont {van Loock}},
  \bibinfo {author} {\bibfnamefont {K.}~\bibnamefont {Nemoto}}, \bibinfo
  {author} {\bibfnamefont {W.~J.}\ \bibnamefont {Munro}}, \ and\ \bibinfo
  {author} {\bibfnamefont {Y.}~\bibnamefont {Yamamoto}},\ }\href {\doibase
  10.1088/1367-2630/8/9/184} {\bibfield  {journal} {\bibinfo  {journal} {New
  Journal of Physics}\ }\textbf {\bibinfo {volume} {8}},\ \bibinfo {pages}
  {184} (\bibinfo {year} {2006})}\BibitemShut {NoStop}%
\bibitem [{\citenamefont {Hartmann}\ \emph {et~al.}(2007)\citenamefont
  {Hartmann}, \citenamefont {Kraus}, \citenamefont {Briegel},\ and\
  \citenamefont {D\"ur}}]{Hartmann_2007}%
  \BibitemOpen
  \bibfield  {author} {\bibinfo {author} {\bibfnamefont {L.}~\bibnamefont
  {Hartmann}}, \bibinfo {author} {\bibfnamefont {B.}~\bibnamefont {Kraus}},
  \bibinfo {author} {\bibfnamefont {H.-J.}\ \bibnamefont {Briegel}}, \ and\
  \bibinfo {author} {\bibfnamefont {W.}~\bibnamefont {D\"ur}},\ }\href
  {\doibase 10.1103/PhysRevA.75.032310} {\bibfield  {journal} {\bibinfo
  {journal} {Phys. Rev. A}\ }\textbf {\bibinfo {volume} {75}},\ \bibinfo
  {pages} {032310} (\bibinfo {year} {2007})}\BibitemShut {NoStop}%
\bibitem [{\citenamefont {Sangouard}\ \emph {et~al.}(2011)\citenamefont
  {Sangouard}, \citenamefont {Simon}, \citenamefont {de~Riedmatten},\ and\
  \citenamefont {Gisin}}]{Sangouard_2011}%
  \BibitemOpen
  \bibfield  {author} {\bibinfo {author} {\bibfnamefont {N.}~\bibnamefont
  {Sangouard}}, \bibinfo {author} {\bibfnamefont {C.}~\bibnamefont {Simon}},
  \bibinfo {author} {\bibfnamefont {H.}~\bibnamefont {de~Riedmatten}}, \ and\
  \bibinfo {author} {\bibfnamefont {N.}~\bibnamefont {Gisin}},\ }\href
  {\doibase 10.1103/RevModPhys.83.33} {\bibfield  {journal} {\bibinfo
  {journal} {Rev. Mod. Phys.}\ }\textbf {\bibinfo {volume} {83}},\ \bibinfo
  {pages} {33} (\bibinfo {year} {2011})}\BibitemShut {NoStop}%
\bibitem [{\citenamefont {Azuma}\ \emph {et~al.}(2015)\citenamefont {Azuma},
  \citenamefont {Tamaki},\ and\ \citenamefont {Lo}}]{Azuma_2015}%
  \BibitemOpen
  \bibfield  {author} {\bibinfo {author} {\bibfnamefont {K.}~\bibnamefont
  {Azuma}}, \bibinfo {author} {\bibfnamefont {K.}~\bibnamefont {Tamaki}}, \
  and\ \bibinfo {author} {\bibfnamefont {H.-K.}\ \bibnamefont {Lo}},\ }\href
  {\doibase 10.1038/ncomms7787} {\bibfield  {journal} {\bibinfo  {journal}
  {Nature Communications}\ }\textbf {\bibinfo {volume} {6}},\ \bibinfo {pages}
  {6787} (\bibinfo {year} {2015})}\BibitemShut {NoStop}%
\bibitem [{\citenamefont {Pirandola}(2016)}]{Pirandola_2016}%
  \BibitemOpen
  \bibfield  {author} {\bibinfo {author} {\bibfnamefont {S.}~\bibnamefont
  {Pirandola}},\ }\href {https://api.semanticscholar.org/CorpusID:119307963}
  {\bibfield  {journal} {\bibinfo  {journal} {arXiv: Quantum Physics}\ }
  (\bibinfo {year} {2016})}\BibitemShut {NoStop}%
\bibitem [{\citenamefont {Azuma}\ and\ \citenamefont
  {Kato}(2017)}]{Azuma_2017}%
  \BibitemOpen
  \bibfield  {author} {\bibinfo {author} {\bibfnamefont {K.}~\bibnamefont
  {Azuma}}\ and\ \bibinfo {author} {\bibfnamefont {G.}~\bibnamefont {Kato}},\
  }\href {\doibase 10.1103/PhysRevA.96.032332} {\bibfield  {journal} {\bibinfo
  {journal} {Phys. Rev. A}\ }\textbf {\bibinfo {volume} {96}},\ \bibinfo
  {pages} {032332} (\bibinfo {year} {2017})}\BibitemShut {NoStop}%
\bibitem [{\citenamefont {Caleffi}(2017)}]{Caleffi_2017}%
  \BibitemOpen
  \bibfield  {author} {\bibinfo {author} {\bibfnamefont {M.}~\bibnamefont
  {Caleffi}},\ }\href {\doibase 10.1109/ACCESS.2017.2763325} {\bibfield
  {journal} {\bibinfo  {journal} {IEEE Access}\ }\textbf {\bibinfo {volume}
  {5}},\ \bibinfo {pages} {22299} (\bibinfo {year} {2017})}\BibitemShut
  {NoStop}%
\bibitem [{\citenamefont {Zwerger}\ \emph {et~al.}(2018)\citenamefont
  {Zwerger}, \citenamefont {Pirker}, \citenamefont {Dunjko}, \citenamefont
  {Briegel},\ and\ \citenamefont {D\"ur}}]{zwerger_18}%
  \BibitemOpen
  \bibfield  {author} {\bibinfo {author} {\bibfnamefont {M.}~\bibnamefont
  {Zwerger}}, \bibinfo {author} {\bibfnamefont {A.}~\bibnamefont {Pirker}},
  \bibinfo {author} {\bibfnamefont {V.}~\bibnamefont {Dunjko}}, \bibinfo
  {author} {\bibfnamefont {H.~J.}\ \bibnamefont {Briegel}}, \ and\ \bibinfo
  {author} {\bibfnamefont {W.}~\bibnamefont {D\"ur}},\ }\href {\doibase
  10.1103/PhysRevLett.120.030503} {\bibfield  {journal} {\bibinfo  {journal}
  {Phys. Rev. Lett.}\ }\textbf {\bibinfo {volume} {120}},\ \bibinfo {pages}
  {030503} (\bibinfo {year} {2018})}\BibitemShut {NoStop}%
\bibitem [{\citenamefont {Banerjee}\ \emph {et~al.}(2020)\citenamefont
  {Banerjee}, \citenamefont {Ghosh}, \citenamefont {Mal},\ and\ \citenamefont
  {Sen(De)}}]{Ratul_2020}%
  \BibitemOpen
  \bibfield  {author} {\bibinfo {author} {\bibfnamefont {R.}~\bibnamefont
  {Banerjee}}, \bibinfo {author} {\bibfnamefont {S.}~\bibnamefont {Ghosh}},
  \bibinfo {author} {\bibfnamefont {S.}~\bibnamefont {Mal}}, \ and\ \bibinfo
  {author} {\bibfnamefont {A.}~\bibnamefont {Sen(De)}},\ }\href {\doibase
  10.1103/PhysRevResearch.2.043355} {\bibfield  {journal} {\bibinfo  {journal}
  {Phys. Rev. Res.}\ }\textbf {\bibinfo {volume} {2}},\ \bibinfo {pages}
  {043355} (\bibinfo {year} {2020})}\BibitemShut {NoStop}%
\bibitem [{\citenamefont {Bugalho}\ \emph {et~al.}(2023)\citenamefont
  {Bugalho}, \citenamefont {Coutinho}, \citenamefont {Monteiro},\ and\
  \citenamefont {Omar}}]{Bugalho_2023}%
  \BibitemOpen
  \bibfield  {author} {\bibinfo {author} {\bibfnamefont {L.}~\bibnamefont
  {Bugalho}}, \bibinfo {author} {\bibfnamefont {B.~C.}\ \bibnamefont
  {Coutinho}}, \bibinfo {author} {\bibfnamefont {F.~A.}\ \bibnamefont
  {Monteiro}}, \ and\ \bibinfo {author} {\bibfnamefont {Y.}~\bibnamefont
  {Omar}},\ }\href {\doibase 10.22331/q-2023-02-09-920} {\bibfield  {journal}
  {\bibinfo  {journal} {{Quantum}}\ }\textbf {\bibinfo {volume} {7}},\ \bibinfo
  {pages} {920} (\bibinfo {year} {2023})}\BibitemShut {NoStop}%
\bibitem [{\citenamefont {Sen(De)}\ \emph {et~al.}(2005)\citenamefont
  {Sen(De)}, \citenamefont {Sen}, \citenamefont {Brukner}, \citenamefont
  {Bu\ifmmode~\check{z}\else \v{z}\fi{}ek},\ and\ \citenamefont
  {\ifmmode~\dot{Z}\else \.{Z}\fi{}ukowski}}]{Aditi_2005}%
  \BibitemOpen
  \bibfield  {author} {\bibinfo {author} {\bibfnamefont {A.}~\bibnamefont
  {Sen(De)}}, \bibinfo {author} {\bibfnamefont {U.}~\bibnamefont {Sen}},
  \bibinfo {author} {\bibfnamefont {C.}~\bibnamefont {Brukner}}, \bibinfo
  {author} {\bibfnamefont {V.}~\bibnamefont {Bu\ifmmode~\check{z}\else
  \v{z}\fi{}ek}}, \ and\ \bibinfo {author} {\bibfnamefont {M.}~\bibnamefont
  {\ifmmode~\dot{Z}\else \.{Z}\fi{}ukowski}},\ }\href {\doibase
  10.1103/PhysRevA.72.042310} {\bibfield  {journal} {\bibinfo  {journal} {Phys.
  Rev. A}\ }\textbf {\bibinfo {volume} {72}},\ \bibinfo {pages} {042310}
  (\bibinfo {year} {2005})}\BibitemShut {NoStop}%
\bibitem [{\citenamefont {Cavalcanti}\ \emph {et~al.}(2011)\citenamefont
  {Cavalcanti}, \citenamefont {Almeida}, \citenamefont {Scarani},\ and\
  \citenamefont {Acín}}]{Cavalcanti_2011}%
  \BibitemOpen
  \bibfield  {author} {\bibinfo {author} {\bibfnamefont {D.}~\bibnamefont
  {Cavalcanti}}, \bibinfo {author} {\bibfnamefont {M.~L.}\ \bibnamefont
  {Almeida}}, \bibinfo {author} {\bibfnamefont {V.}~\bibnamefont {Scarani}}, \
  and\ \bibinfo {author} {\bibfnamefont {A.}~\bibnamefont {Acín}},\ }\href
  {\doibase 10.1038/ncomms1193} {\bibfield  {journal} {\bibinfo  {journal}
  {Nature Communications}\ }\textbf {\bibinfo {volume} {2}},\ \bibinfo {pages}
  {184} (\bibinfo {year} {2011})}\BibitemShut {NoStop}%
\bibitem [{\citenamefont {Lapeyre}\ \emph {et~al.}(2009)\citenamefont
  {Lapeyre}, \citenamefont {Wehr},\ and\ \citenamefont
  {Lewenstein}}]{lapeyre09}%
  \BibitemOpen
  \bibfield  {author} {\bibinfo {author} {\bibfnamefont {G.~J.}\ \bibnamefont
  {Lapeyre}}, \bibinfo {author} {\bibfnamefont {J.}~\bibnamefont {Wehr}}, \
  and\ \bibinfo {author} {\bibfnamefont {M.}~\bibnamefont {Lewenstein}},\
  }\href {\doibase 10.1103/PhysRevA.79.042324} {\bibfield  {journal} {\bibinfo
  {journal} {Phys. Rev. A}\ }\textbf {\bibinfo {volume} {79}},\ \bibinfo
  {pages} {042324} (\bibinfo {year} {2009})}\BibitemShut {NoStop}%
\bibitem [{\citenamefont {Cuquet}\ and\ \citenamefont
  {Calsamiglia}(2009)}]{cuquet09}%
  \BibitemOpen
  \bibfield  {author} {\bibinfo {author} {\bibfnamefont {M.}~\bibnamefont
  {Cuquet}}\ and\ \bibinfo {author} {\bibfnamefont {J.}~\bibnamefont
  {Calsamiglia}},\ }\href {\doibase 10.1103/PhysRevLett.103.240503} {\bibfield
  {journal} {\bibinfo  {journal} {Phys. Rev. Lett.}\ }\textbf {\bibinfo
  {volume} {103}},\ \bibinfo {pages} {240503} (\bibinfo {year}
  {2009})}\BibitemShut {NoStop}%
\bibitem [{\citenamefont {Broadfoot}\ \emph {et~al.}(2009)\citenamefont
  {Broadfoot}, \citenamefont {Dorner},\ and\ \citenamefont
  {Jaksch}}]{Broadfoot2009}%
  \BibitemOpen
  \bibfield  {author} {\bibinfo {author} {\bibfnamefont {S.}~\bibnamefont
  {Broadfoot}}, \bibinfo {author} {\bibfnamefont {U.}~\bibnamefont {Dorner}}, \
  and\ \bibinfo {author} {\bibfnamefont {D.}~\bibnamefont {Jaksch}},\ }\href
  {\doibase 10.1209/0295-5075/88/50002} {\bibfield  {journal} {\bibinfo
  {journal} {EPL (Europhysics Letters)}\ }\textbf {\bibinfo {volume} {88}},\
  \bibinfo {pages} {50002} (\bibinfo {year} {2009})}\BibitemShut {NoStop}%
\bibitem [{\citenamefont {Perseguers}\ \emph {et~al.}(2010)\citenamefont
  {Perseguers}, \citenamefont {Cavalcanti}, \citenamefont {Lapeyre},
  \citenamefont {Lewenstein},\ and\ \citenamefont {Ac\'{\i}n}}]{perseguers10}%
  \BibitemOpen
  \bibfield  {author} {\bibinfo {author} {\bibfnamefont {S.}~\bibnamefont
  {Perseguers}}, \bibinfo {author} {\bibfnamefont {D.}~\bibnamefont
  {Cavalcanti}}, \bibinfo {author} {\bibfnamefont {G.~J.}\ \bibnamefont
  {Lapeyre}}, \bibinfo {author} {\bibfnamefont {M.}~\bibnamefont {Lewenstein}},
  \ and\ \bibinfo {author} {\bibfnamefont {A.}~\bibnamefont {Ac\'{\i}n}},\
  }\href {\doibase 10.1103/PhysRevA.81.032327} {\bibfield  {journal} {\bibinfo
  {journal} {Phys. Rev. A}\ }\textbf {\bibinfo {volume} {81}},\ \bibinfo
  {pages} {032327} (\bibinfo {year} {2010})}\BibitemShut {NoStop}%
\bibitem [{\citenamefont {Lee}\ and\ \citenamefont {Khitrin}(2005)}]{Lee_2005}%
  \BibitemOpen
  \bibfield  {author} {\bibinfo {author} {\bibfnamefont {J.-S.}\ \bibnamefont
  {Lee}}\ and\ \bibinfo {author} {\bibfnamefont {A.~K.}\ \bibnamefont
  {Khitrin}},\ }\href {\doibase 10.1103/PhysRevLett.94.150504} {\bibfield
  {journal} {\bibinfo  {journal} {Phys. Rev. Lett.}\ }\textbf {\bibinfo
  {volume} {94}},\ \bibinfo {pages} {150504} (\bibinfo {year}
  {2005})}\BibitemShut {NoStop}%
\bibitem [{\citenamefont {\"Ozdemir}\ \emph {et~al.}(2011)\citenamefont
  {\"Ozdemir}, \citenamefont {Matsunaga}, \citenamefont {Tashima},
  \citenamefont {Yamamoto}, \citenamefont {Koashi},\ and\ \citenamefont
  {Imoto}}]{Ozdemir_2011}%
  \BibitemOpen
  \bibfield  {author} {\bibinfo {author} {\bibfnamefont {S.~K.}\ \bibnamefont
  {\"Ozdemir}}, \bibinfo {author} {\bibfnamefont {E.}~\bibnamefont
  {Matsunaga}}, \bibinfo {author} {\bibfnamefont {T.}~\bibnamefont {Tashima}},
  \bibinfo {author} {\bibfnamefont {T.}~\bibnamefont {Yamamoto}}, \bibinfo
  {author} {\bibfnamefont {M.}~\bibnamefont {Koashi}}, \ and\ \bibinfo {author}
  {\bibfnamefont {N.}~\bibnamefont {Imoto}},\ }\href {\doibase
  10.1088/1367-2630/13/10/103003} {\bibfield  {journal} {\bibinfo  {journal}
  {New Journal of Physics}\ }\textbf {\bibinfo {volume} {13}},\ \bibinfo
  {pages} {103003} (\bibinfo {year} {2011})}\BibitemShut {NoStop}%
\bibitem [{\citenamefont {Zang}\ \emph {et~al.}(2015)\citenamefont {Zang},
  \citenamefont {Yang}, \citenamefont {Ozaydin}, \citenamefont {Song},\ and\
  \citenamefont {Cao}}]{Zang_2015}%
  \BibitemOpen
  \bibfield  {author} {\bibinfo {author} {\bibfnamefont {X.-P.}\ \bibnamefont
  {Zang}}, \bibinfo {author} {\bibfnamefont {M.}~\bibnamefont {Yang}}, \bibinfo
  {author} {\bibfnamefont {F.}~\bibnamefont {Ozaydin}}, \bibinfo {author}
  {\bibfnamefont {W.}~\bibnamefont {Song}}, \ and\ \bibinfo {author}
  {\bibfnamefont {Z.-L.}\ \bibnamefont {Cao}},\ }\href {\doibase
  10.1038/srep16245} {\bibfield  {journal} {\bibinfo  {journal} {Scientific
  Reports}\ }\textbf {\bibinfo {volume} {5}},\ \bibinfo {pages} {16245}
  (\bibinfo {year} {2015})}\BibitemShut {NoStop}%
\bibitem [{\citenamefont {Halder}\ \emph
  {et~al.}(2022{\natexlab{a}})\citenamefont {Halder}, \citenamefont {Banerjee},
  \citenamefont {Ghosh}, \citenamefont {Pal},\ and\ \citenamefont
  {Sen(De)}}]{Pritam_2022}%
  \BibitemOpen
  \bibfield  {author} {\bibinfo {author} {\bibfnamefont {P.}~\bibnamefont
  {Halder}}, \bibinfo {author} {\bibfnamefont {R.}~\bibnamefont {Banerjee}},
  \bibinfo {author} {\bibfnamefont {S.}~\bibnamefont {Ghosh}}, \bibinfo
  {author} {\bibfnamefont {A.~K.}\ \bibnamefont {Pal}}, \ and\ \bibinfo
  {author} {\bibfnamefont {A.}~\bibnamefont {Sen(De)}},\ }\href {\doibase
  10.1103/PhysRevA.106.032604} {\bibfield  {journal} {\bibinfo  {journal}
  {Phys. Rev. A}\ }\textbf {\bibinfo {volume} {106}},\ \bibinfo {pages}
  {032604} (\bibinfo {year} {2022}{\natexlab{a}})}\BibitemShut {NoStop}%
\bibitem [{\citenamefont {Busch}\ \emph {et~al.}(2013)\citenamefont {Busch},
  \citenamefont {Lahti},\ and\ \citenamefont {Mittelstaedt}}]{Busch_2013}%
  \BibitemOpen
  \bibfield  {author} {\bibinfo {author} {\bibfnamefont {P.}~\bibnamefont
  {Busch}}, \bibinfo {author} {\bibfnamefont {P.~J.}\ \bibnamefont {Lahti}}, \
  and\ \bibinfo {author} {\bibfnamefont {P.}~\bibnamefont {Mittelstaedt}},\
  }\href@noop {} {\emph {\bibinfo {title} {The Quantum Theory of
  Measurement}}},\ Vol.~\bibinfo {volume} {2}\ (\bibinfo {year}
  {2013})\BibitemShut {NoStop}%
\bibitem [{\citenamefont {Mal}\ \emph {et~al.}(2016)\citenamefont {Mal},
  \citenamefont {Majumdar},\ and\ \citenamefont {Home}}]{Mal_2016}%
  \BibitemOpen
  \bibfield  {author} {\bibinfo {author} {\bibfnamefont {S.}~\bibnamefont
  {Mal}}, \bibinfo {author} {\bibfnamefont {A.~S.}\ \bibnamefont {Majumdar}}, \
  and\ \bibinfo {author} {\bibfnamefont {D.}~\bibnamefont {Home}},\ }\href
  {\doibase 10.3390/math4030048} {\bibfield  {journal} {\bibinfo  {journal}
  {Mathematics}\ }\textbf {\bibinfo {volume} {4}} (\bibinfo {year} {2016}),\
  10.3390/math4030048}\BibitemShut {NoStop}%
\bibitem [{\citenamefont {Srivastava}\ \emph {et~al.}(2021)\citenamefont
  {Srivastava}, \citenamefont {Mal}, \citenamefont {Sen(De)},\ and\
  \citenamefont {Sen}}]{srivastava21}%
  \BibitemOpen
  \bibfield  {author} {\bibinfo {author} {\bibfnamefont {C.}~\bibnamefont
  {Srivastava}}, \bibinfo {author} {\bibfnamefont {S.}~\bibnamefont {Mal}},
  \bibinfo {author} {\bibfnamefont {A.}~\bibnamefont {Sen(De)}}, \ and\
  \bibinfo {author} {\bibfnamefont {U.}~\bibnamefont {Sen}},\ }\href {\doibase
  10.1103/PhysRevA.103.032408} {\bibfield  {journal} {\bibinfo  {journal}
  {Phys. Rev. A}\ }\textbf {\bibinfo {volume} {103}},\ \bibinfo {pages}
  {032408} (\bibinfo {year} {2021})}\BibitemShut {NoStop}%
\bibitem [{\citenamefont {Roy}\ \emph {et~al.}(2021)\citenamefont {Roy},
  \citenamefont {Bera}, \citenamefont {Mal}, \citenamefont {Sen(De)},\ and\
  \citenamefont {Sen}}]{ROY2021127143}%
  \BibitemOpen
  \bibfield  {author} {\bibinfo {author} {\bibfnamefont {S.}~\bibnamefont
  {Roy}}, \bibinfo {author} {\bibfnamefont {A.}~\bibnamefont {Bera}}, \bibinfo
  {author} {\bibfnamefont {S.}~\bibnamefont {Mal}}, \bibinfo {author}
  {\bibfnamefont {A.}~\bibnamefont {Sen(De)}}, \ and\ \bibinfo {author}
  {\bibfnamefont {U.}~\bibnamefont {Sen}},\ }\href {\doibase
  https://doi.org/10.1016/j.physleta.2021.127143} {\bibfield  {journal}
  {\bibinfo  {journal} {Physics Letters A}\ }\textbf {\bibinfo {volume}
  {392}},\ \bibinfo {pages} {127143} (\bibinfo {year} {2021})}\BibitemShut
  {NoStop}%
\bibitem [{\citenamefont {Lv}\ \emph {et~al.}(2023)\citenamefont {Lv},
  \citenamefont {Liang}, \citenamefont {Wang},\ and\ \citenamefont
  {Fei}}]{Lv2023}%
  \BibitemOpen
  \bibfield  {author} {\bibinfo {author} {\bibfnamefont {Q.-Q.}\ \bibnamefont
  {Lv}}, \bibinfo {author} {\bibfnamefont {J.-M.}\ \bibnamefont {Liang}},
  \bibinfo {author} {\bibfnamefont {Z.-X.}\ \bibnamefont {Wang}}, \ and\
  \bibinfo {author} {\bibfnamefont {S.-M.}\ \bibnamefont {Fei}},\ }\href
  {\doibase 10.1088/1751-8121/ace504} {\bibfield  {journal} {\bibinfo
  {journal} {Journal of Physics A: Mathematical and Theoretical}\ }\textbf
  {\bibinfo {volume} {56}},\ \bibinfo {pages} {325301} (\bibinfo {year}
  {2023})}\BibitemShut {NoStop}%
\bibitem [{\citenamefont {Halder}\ \emph
  {et~al.}(2022{\natexlab{b}})\citenamefont {Halder}, \citenamefont {Banerjee},
  \citenamefont {Mal},\ and\ \citenamefont {Sen(De)}}]{Halder_2022}%
  \BibitemOpen
  \bibfield  {author} {\bibinfo {author} {\bibfnamefont {P.}~\bibnamefont
  {Halder}}, \bibinfo {author} {\bibfnamefont {R.}~\bibnamefont {Banerjee}},
  \bibinfo {author} {\bibfnamefont {S.}~\bibnamefont {Mal}}, \ and\ \bibinfo
  {author} {\bibfnamefont {A.}~\bibnamefont {Sen(De)}},\ }\href {\doibase
  10.1103/PhysRevA.106.052413} {\bibfield  {journal} {\bibinfo  {journal}
  {Phys. Rev. A}\ }\textbf {\bibinfo {volume} {106}},\ \bibinfo {pages}
  {052413} (\bibinfo {year} {2022}{\natexlab{b}})}\BibitemShut {NoStop}%
\bibitem [{\citenamefont {Zhang}\ \emph
  {et~al.}(2023{\natexlab{a}})\citenamefont {Zhang}, \citenamefont {Jing},\
  and\ \citenamefont {Fei}}]{zhang_2023}%
  \BibitemOpen
  \bibfield  {author} {\bibinfo {author} {\bibfnamefont {T.}~\bibnamefont
  {Zhang}}, \bibinfo {author} {\bibfnamefont {N.}~\bibnamefont {Jing}}, \ and\
  \bibinfo {author} {\bibfnamefont {S.-M.}\ \bibnamefont {Fei}},\ }\href
  {\doibase 10.1007/s11467-022-1242-6} {\bibfield  {journal} {\bibinfo
  {journal} {Frontiers of Physics}\ }\textbf {\bibinfo {volume} {18}},\
  \bibinfo {pages} {31302} (\bibinfo {year} {2023}{\natexlab{a}})}\BibitemShut
  {NoStop}%
\bibitem [{\citenamefont {Halder}\ \emph {et~al.}(2021)\citenamefont {Halder},
  \citenamefont {Mal},\ and\ \citenamefont {Sen(De)}}]{inflation21}%
  \BibitemOpen
  \bibfield  {author} {\bibinfo {author} {\bibfnamefont {P.}~\bibnamefont
  {Halder}}, \bibinfo {author} {\bibfnamefont {S.}~\bibnamefont {Mal}}, \ and\
  \bibinfo {author} {\bibfnamefont {A.}~\bibnamefont {Sen(De)}},\ }\href
  {\doibase 10.1103/PhysRevA.104.062412} {\bibfield  {journal} {\bibinfo
  {journal} {Phys. Rev. A}\ }\textbf {\bibinfo {volume} {104}},\ \bibinfo
  {pages} {062412} (\bibinfo {year} {2021})}\BibitemShut {NoStop}%
\bibitem [{\citenamefont {Coffman}\ \emph {et~al.}(2000)\citenamefont
  {Coffman}, \citenamefont {Kundu},\ and\ \citenamefont {Wootters}}]{ckw00}%
  \BibitemOpen
  \bibfield  {author} {\bibinfo {author} {\bibfnamefont {V.}~\bibnamefont
  {Coffman}}, \bibinfo {author} {\bibfnamefont {J.}~\bibnamefont {Kundu}}, \
  and\ \bibinfo {author} {\bibfnamefont {W.~K.}\ \bibnamefont {Wootters}},\
  }\href {\doibase 10.1103/PhysRevA.61.052306} {\bibfield  {journal} {\bibinfo
  {journal} {Phys. Rev. A}\ }\textbf {\bibinfo {volume} {61}},\ \bibinfo
  {pages} {052306} (\bibinfo {year} {2000})}\BibitemShut {NoStop}%
\bibitem [{\citenamefont {Ou}\ and\ \citenamefont {Fan}(2007)}]{ou07}%
  \BibitemOpen
  \bibfield  {author} {\bibinfo {author} {\bibfnamefont {Y.-C.}\ \bibnamefont
  {Ou}}\ and\ \bibinfo {author} {\bibfnamefont {H.}~\bibnamefont {Fan}},\
  }\href {\doibase 10.1103/PhysRevA.75.062308} {\bibfield  {journal} {\bibinfo
  {journal} {Phys. Rev. A}\ }\textbf {\bibinfo {volume} {75}},\ \bibinfo
  {pages} {062308} (\bibinfo {year} {2007})}\BibitemShut {NoStop}%
\bibitem [{\citenamefont {Dhar}\ \emph {et~al.}(2017)\citenamefont {Dhar},
  \citenamefont {Pal}, \citenamefont {Rakshit}, \citenamefont {Sen(De)},\ and\
  \citenamefont {Sen}}]{monorev17}%
  \BibitemOpen
  \bibfield  {author} {\bibinfo {author} {\bibfnamefont {H.~S.}\ \bibnamefont
  {Dhar}}, \bibinfo {author} {\bibfnamefont {A.~K.}\ \bibnamefont {Pal}},
  \bibinfo {author} {\bibfnamefont {D.}~\bibnamefont {Rakshit}}, \bibinfo
  {author} {\bibfnamefont {A.}~\bibnamefont {Sen(De)}}, \ and\ \bibinfo
  {author} {\bibfnamefont {U.}~\bibnamefont {Sen}},\ }in\ \href {\doibase
  10.1007/978-3-319-53412-1_3} {\emph {\bibinfo {booktitle} {Quantum Science
  and Technology}}}\ (\bibinfo  {publisher} {Springer International
  Publishing},\ \bibinfo {year} {2017})\ pp.\ \bibinfo {pages}
  {23--64}\BibitemShut {NoStop}%
\bibitem [{\citenamefont {Sen(De)}\ and\ \citenamefont
  {Sen}(2010{\natexlab{b}})}]{ggm10}%
  \BibitemOpen
  \bibfield  {author} {\bibinfo {author} {\bibfnamefont {A.}~\bibnamefont
  {Sen(De)}}\ and\ \bibinfo {author} {\bibfnamefont {U.}~\bibnamefont {Sen}},\
  }\href {\doibase 10.1103/PhysRevA.81.012308} {\bibfield  {journal} {\bibinfo
  {journal} {Phys. Rev. A}\ }\textbf {\bibinfo {volume} {81}},\ \bibinfo
  {pages} {012308} (\bibinfo {year} {2010}{\natexlab{b}})}\BibitemShut
  {NoStop}%
\bibitem [{\citenamefont {Biswas}\ \emph {et~al.}(2014)\citenamefont {Biswas},
  \citenamefont {Prabhu}, \citenamefont {Sen(De)},\ and\ \citenamefont
  {Sen}}]{ggm14}%
  \BibitemOpen
  \bibfield  {author} {\bibinfo {author} {\bibfnamefont {A.}~\bibnamefont
  {Biswas}}, \bibinfo {author} {\bibfnamefont {R.}~\bibnamefont {Prabhu}},
  \bibinfo {author} {\bibfnamefont {A.}~\bibnamefont {Sen(De)}}, \ and\
  \bibinfo {author} {\bibfnamefont {U.}~\bibnamefont {Sen}},\ }\href {\doibase
  10.1103/PhysRevA.90.032301} {\bibfield  {journal} {\bibinfo  {journal} {Phys.
  Rev. A}\ }\textbf {\bibinfo {volume} {90}},\ \bibinfo {pages} {032301}
  (\bibinfo {year} {2014})}\BibitemShut {NoStop}%
\bibitem [{\citenamefont {Das}\ \emph {et~al.}(2016)\citenamefont {Das},
  \citenamefont {Roy}, \citenamefont {Bagchi}, \citenamefont {Misra},
  \citenamefont {Sen(De)},\ and\ \citenamefont {Sen}}]{tamoghna16mixed_ggm}%
  \BibitemOpen
  \bibfield  {author} {\bibinfo {author} {\bibfnamefont {T.}~\bibnamefont
  {Das}}, \bibinfo {author} {\bibfnamefont {S.~S.}\ \bibnamefont {Roy}},
  \bibinfo {author} {\bibfnamefont {S.}~\bibnamefont {Bagchi}}, \bibinfo
  {author} {\bibfnamefont {A.}~\bibnamefont {Misra}}, \bibinfo {author}
  {\bibfnamefont {A.}~\bibnamefont {Sen(De)}}, \ and\ \bibinfo {author}
  {\bibfnamefont {U.}~\bibnamefont {Sen}},\ }\href {\doibase
  10.1103/PhysRevA.94.022336} {\bibfield  {journal} {\bibinfo  {journal} {Phys.
  Rev. A}\ }\textbf {\bibinfo {volume} {94}},\ \bibinfo {pages} {022336}
  (\bibinfo {year} {2016})}\BibitemShut {NoStop}%
\bibitem [{\citenamefont {Buchholz}\ \emph {et~al.}(2016)\citenamefont
  {Buchholz}, \citenamefont {Moroder},\ and\ \citenamefont
  {Gühne}}]{Buchholz2016}%
  \BibitemOpen
  \bibfield  {author} {\bibinfo {author} {\bibfnamefont {L.~E.}\ \bibnamefont
  {Buchholz}}, \bibinfo {author} {\bibfnamefont {T.}~\bibnamefont {Moroder}}, \
  and\ \bibinfo {author} {\bibfnamefont {O.}~\bibnamefont {Gühne}},\ }\href
  {\doibase 10.1002/andp.201500293} {\bibfield  {journal} {\bibinfo  {journal}
  {Annalen der Physik}\ }\textbf {\bibinfo {volume} {528}},\ \bibinfo {pages}
  {278} (\bibinfo {year} {2016})}\BibitemShut {NoStop}%
\bibitem [{\citenamefont {Osborne}\ and\ \citenamefont
  {Verstraete}(2006)}]{osborne06}%
  \BibitemOpen
  \bibfield  {author} {\bibinfo {author} {\bibfnamefont {T.~J.}\ \bibnamefont
  {Osborne}}\ and\ \bibinfo {author} {\bibfnamefont {F.}~\bibnamefont
  {Verstraete}},\ }\href {\doibase 10.1103/PhysRevLett.96.220503} {\bibfield
  {journal} {\bibinfo  {journal} {Phys. Rev. Lett.}\ }\textbf {\bibinfo
  {volume} {96}},\ \bibinfo {pages} {220503} (\bibinfo {year}
  {2006})}\BibitemShut {NoStop}%
\bibitem [{\citenamefont {Zhang}\ \emph
  {et~al.}(2023{\natexlab{b}})\citenamefont {Zhang}, \citenamefont {Jing},
  \citenamefont {Liu},\ and\ \citenamefont {Ma}}]{Zhang2023}%
  \BibitemOpen
  \bibfield  {author} {\bibinfo {author} {\bibfnamefont {X.}~\bibnamefont
  {Zhang}}, \bibinfo {author} {\bibfnamefont {N.}~\bibnamefont {Jing}},
  \bibinfo {author} {\bibfnamefont {M.}~\bibnamefont {Liu}}, \ and\ \bibinfo
  {author} {\bibfnamefont {H.}~\bibnamefont {Ma}},\ }\href {\doibase
  10.1088/1402-4896/acbb37} {\bibfield  {journal} {\bibinfo  {journal} {Physica
  Scripta}\ }\textbf {\bibinfo {volume} {98}},\ \bibinfo {pages} {035106}
  (\bibinfo {year} {2023}{\natexlab{b}})}\BibitemShut {NoStop}%
\bibitem [{\citenamefont {Vidal}\ and\ \citenamefont
  {Werner}(2002)}]{vidal02neg}%
  \BibitemOpen
  \bibfield  {author} {\bibinfo {author} {\bibfnamefont {G.}~\bibnamefont
  {Vidal}}\ and\ \bibinfo {author} {\bibfnamefont {R.~F.}\ \bibnamefont
  {Werner}},\ }\href {\doibase 10.1103/PhysRevA.65.032314} {\bibfield
  {journal} {\bibinfo  {journal} {Phys. Rev. A}\ }\textbf {\bibinfo {volume}
  {65}},\ \bibinfo {pages} {032314} (\bibinfo {year} {2002})}\BibitemShut
  {NoStop}%
\bibitem [{\citenamefont {Plenio}(2005)}]{plenio05logneg}%
  \BibitemOpen
  \bibfield  {author} {\bibinfo {author} {\bibfnamefont {M.~B.}\ \bibnamefont
  {Plenio}},\ }\href {\doibase 10.1103/PhysRevLett.95.090503} {\bibfield
  {journal} {\bibinfo  {journal} {Phys. Rev. Lett.}\ }\textbf {\bibinfo
  {volume} {95}},\ \bibinfo {pages} {090503} (\bibinfo {year}
  {2005})}\BibitemShut {NoStop}%
\bibitem [{\citenamefont {Peres}(1996)}]{peres96ppt}%
  \BibitemOpen
  \bibfield  {author} {\bibinfo {author} {\bibfnamefont {A.}~\bibnamefont
  {Peres}},\ }\href {\doibase 10.1103/PhysRevLett.77.1413} {\bibfield
  {journal} {\bibinfo  {journal} {Phys. Rev. Lett.}\ }\textbf {\bibinfo
  {volume} {77}},\ \bibinfo {pages} {1413} (\bibinfo {year}
  {1996})}\BibitemShut {NoStop}%
\bibitem [{\citenamefont {Horodecki}\ \emph {et~al.}(1996)\citenamefont
  {Horodecki}, \citenamefont {Horodecki},\ and\ \citenamefont
  {Horodecki}}]{Horodecki1996}%
  \BibitemOpen
  \bibfield  {author} {\bibinfo {author} {\bibfnamefont {M.}~\bibnamefont
  {Horodecki}}, \bibinfo {author} {\bibfnamefont {P.}~\bibnamefont
  {Horodecki}}, \ and\ \bibinfo {author} {\bibfnamefont {R.}~\bibnamefont
  {Horodecki}},\ }\href {\doibase
  https://doi.org/10.1016/S0375-9601(96)00706-2} {\bibfield  {journal}
  {\bibinfo  {journal} {Physics Letters A}\ }\textbf {\bibinfo {volume}
  {223}},\ \bibinfo {pages} {1} (\bibinfo {year} {1996})}\BibitemShut {NoStop}%
\bibitem [{\citenamefont {Bengtsson}\ and\ \citenamefont
  {Zyczkowski}(2006)}]{Bengtsson2006}%
  \BibitemOpen
  \bibfield  {author} {\bibinfo {author} {\bibfnamefont {I.}~\bibnamefont
  {Bengtsson}}\ and\ \bibinfo {author} {\bibfnamefont {K.}~\bibnamefont
  {Zyczkowski}},\ }\href {\doibase 10.1017/CBO9780511535048} {\emph {\bibinfo
  {title} {Geometry of Quantum States}}}\ (\bibinfo  {publisher} {Cambridge
  University Press},\ \bibinfo {year} {2006})\BibitemShut {NoStop}%
\end{thebibliography}%

\end{document}